# Elemental and isotopic variability in solar system materials by mixing and processing of primordial disk reservoirs

Christoph Burkhardt[1,2*], Nicolas Dauphas[2], Ulrik Hans[3], Bernard Bourdon[4] , Thorsten Kleine[1]

[1]Institut für Planetologie, University of Münster, Wilhelm Klemm-Straße 10, D-48149 Münster, Germany
[2]Origins Laboratory, Department of the Geophysical Sciences and Enrico Fermi Institute, The University of Chicago, 5734 South Ellis Avenue, Chicago, Illinois 60637, USA
[3]EMPA, Laboratory for Advanced Materials and Surfaces, Überlandstrasse 129, CH-8600 Dübendorf, Switzerland
[4]Laboratoire de Géologie de Lyon, ENS Lyon, CNRS and Université Claude Bernard Lyon 1, 69364 Lyon cedex 07, France

[*]corresponding author:
  e-mail: burkhardt@uni-muenster.de
  phone: +49 251 83-39039
  fax: +49 251 83-36301

**Abstract**

Isotope anomalies among planetary bodies provide key constraints on planetary genetics and the Solar System's dynamical evolution. However, to unlock the full potential of these anomalies for constraining the processing, mixing, and transport of material in the disk it is essential to identify the main components responsible for producing planetary-scale isotope variations, and to investigate how they relate to the isotopic heterogeneity inherited from the Solar System's parental molecular cloud. To address these issues we measured the Ti and Sr isotopic compositions of Ca,Al-rich inclusions (CAIs) from the Allende CV3 chondrite, as well as acid leachates and an insoluble residue from the Murchison CM2 chondrite, and combine these results with literature data for presolar grains, hibonites, chondrules, and bulk meteorites. Our analysis reveals that the internal mineral-scale nebular isotopic heterogeneity as sampled by leachates and presolar grains is largely decoupled from the planetary-scale isotope anomalies as sampled by bulk meteorites. We show that variable admixing of CAI-like refractory material to an average inner solar nebula component can explain the planetary-scale Ti and Sr isotope anomalies and the elemental and isotopic difference between non-carbonaceous (NC) and carbonaceous (CC) nebular reservoirs for these elements.

Combining isotope anomaly data for a large number of elements (Ti, Sr, Ca, Cr, Ni, Zr, Mo, Ru, Ba, Nd, Sm, Hf, W, and Os) reveals that the offset of the CC from the NC reservoir towards the composition of CAIs is a general trend and not limited to refractory elements. This implies that the CC reservoir is the product of mixing between NC material and a reservoir (called IC for Inclusion-like Chondritic component) whose isotopic composition is similar to that of CAIs, but whose chemical composition is similar to bulk chondrites. In our preferred model, the distinct isotopic compositions of these two nebular reservoirs reflect an inherited heterogeneity of the solar system's parental molecular cloud core, which therefore has never been fully homogenized during collapse. Planetary-scale isotopic anomalies are thus caused by variable mixing of isotopically distinct primordial disk reservoirs, the selective processing of these reservoirs in different nebular environments, and the heterogeneous distribution of the thereby forming nebular products.

## 1. Introduction

The solar system formed ~4.567 Ga ago by the gravitational collapse of a cold and dense molecular cloud core whose chemical and isotopic composition reflects ~9 Ga of galactic chemical evolution (*e.g.*, Dauphas, 2005). Insights into the nature of the materials present at solar system formation—and the processes that acted on them—can be gained by analyzing chondritic meteorites, which derive from asteroids that have not undergone melting and chemical differentiation, and as such have preserved a record of the processes that affected solid material in the circum-stellar disk prior to their incorporation into planetary bodies.

The identification of extreme mass-independent isotopic anomalies in minerals from the matrices of chondrites reveals that the solar system's parental molecular cloud contained dust particles with distinctive nucleosynthetic signatures inherited from previous generations of stars, and that some of this dust survived processing in the solar system (*e.g.*, Zinner 2014). However, these presolar grains make up at most 0.2 ‰ of a primitive chondrite. All other chondrite components show signs of reprocessing in the interstellar medium (*e.g.*, aggregate formation and amorphization by cosmic rays), high-temperature processing in the solar accretion disk (*e.g.*, evaporation, condensation, and melting), and aqueous alteration and heating on meteorite parent bodies. All these processes likely contributed to the homogenization of primary nucleosynthetic signatures. Nevertheless, small but resolvable isotope anomalies are present in most solar system materials (for a recent summary see Dauphas and Schauble 2016).

Different chondrite classes exhibit significant textural, mineralogical, and chemical variability (*e.g.*, in abundance of volatile elements or CAIs, in the ratio of chondrules to matrix, or in the bulk oxygen contents), attesting to the variable physico-chemical conditions that prevailed in the early solar system (Scott and Krot, 2014). These variable conditions provide evidence for selective processing and fractionation of molecular cloud material in different regions of the solar nebula and at different times. Therefore, it is no surprise that resolvable nucleosynthetic isotope variations are present among bulk meteorites. Although the resulting isotopic variations are small (*i.e.*, in the parts-per-ten-thousand to parts-per-million range; compared to orders-of-magnitude variations in presolar grains of circumstellar origin), they provide a means of testing models for the dynamical evolution of the (proto-)planetary disk and the accretion history of planets. For example, nucleosynthetic isotopic anomalies (*i*) affect radioisotope systems used to establish the chronology of the solar system (Burkhardt et al. 2012a; 2016; Boyet et al. 2018; Hans et al. 2013; Render et al. 2017); (*ii*) allow determining the building material of the Earth (*e.g.*, Burkhardt et al., 2011; 2016; Warren, 2011) and how this may have changed over time (Dauphas 2017; Schiller et al. 2018; Budde et al. 2019); and (*iii*) drive the development of new types of giant impact models for the origin of the Moon (*e.g.*, Pahlevan and Stevenson, 2007; Lock et al., 2018).

Finally, in the past few years it has been realized that nucleosynthetic isotope anomalies allow distinguishing between material originating from at least two fundamentally different nebular source regions, which have been termed the non-carbonaceous (NC) and carbonaceous (CC) reservoirs (Warren, 2011; Budde et al., 2016). These two reservoirs have been isolated from each other for several million years, most likely due to the early formation of Jupiter in between them (Budde et al. 2016; Kruijer et al. 2017; Van Kooten et al. 2016; Warren 2011). Although it has been long known that most carbonaceous chondrites are isotopically distinct from other meteoritic and planetary materials (*e.g.*, Clayton et al. 1976; Niemeyer 1988; Leya et al. 2008), and it has been noted that in multi-elemental isotope space meteorites define distinct clusters (Trinquier et al. 2007; 2009; Regelous et al. 2008), it was Warren (2011) who pointed out the fundamental 'gap' between the two clusters and who introduced the terms non-carbonaceous (NC) and carbonaceous meteorites (CC) as taxonomic super-groups. Subsequent work demonstrated that the NC-CC dichotomy extends to a wide range of elements (*e.g.*, Mo, Ni, Ru, W), and also holds for differentiated meteorites, which derive from bodies that accreted within the first ~1 Ma of the solar system. As such, the NC-CC dichotomy is a fundamental and ubiquitous characteristic of the early solar system that has been established very early.

These examples highlight how planetary-scale nucleosynthetic anomalies provide key constraints on fundamental aspects of disk dynamics and planet formation. However, a clear understanding of how processing, mixing, and transport of dust during the early evolution of the solar accretion disk gave rise to large-scale isotopic anomalies is still lacking. Two endmember scenarios have been proposed to account for the isotopic heterogeneity observed at the bulk meteorite scale (Fig. 1). Either the bulk anomalies are a primordial feature of the solar nebula inherited from heterogeneities in the infalling molecular cloud material, *e.g.*, due to a late injection of freshly synthesized matter (Clayton 1982; Dauphas et al. 2002; 2004; Bizzarro et al. 2007), or they are caused by fractionation processes of an initially homogeneous cloud, either during infall (Van Kooten et al. 2016), or within the solar nebula itself (Dauphas et al. 2004; 2010; Trinquier et al. 2009; Burkhardt et al. 2012b; Pignatale et al. 2017; Regelous et al. 2008; Akram et al. 2015; Paton et al. 2013; Schiller et al. 2015; Poole et al. 2017). Such fractionation processes may involve the separation of dust grains of different types and sizes, or the selective destruction of thermally labile presolar components. For example, transport of dust in the disk, settling of dust to the midplane, or turbulences in the disk may have fractionated small presolar grains from larger isotopically more normal dust, leading to the formation of isotopically distinct disk regions (*e.g.*, Dauphas et al. 2010). Likewise, thermal processing of dust along gradients in the disk will affect refractory presolar phases differently than thermally more labile phases, and may also have produced isotopically distinct disk reservoirs (*e.g.*, Burkhardt et al. 2012b). Although these scenarios and processes can account for the presence of ε-level isotopic anomalies in bulk meteorites and planets, until now they have not been linked to specific nebular conditions. As such, a holistic and quantitative model for the distribution of nucleosynthetic anomalies in planetary materials is lacking. This is because for many elements neither the presolar carriers of the nucleosynthetic anomalies, nor their behavior in different nebular settings are known. Furthermore, for many elements the individual presolar carriers may relate to the generation of bulk anomalies only in an indirect way, through the redistribution of materials (such as CAIs and chondrules) formed by dust processing in the solar nebula.

Here we combine new concentration and Ti and Sr isotope data for acid leachates and an insoluble residue from the Murchison CM2 carbonaceous chondrite, as well as for Allende CAIs, with existing isotope anomaly data for nebular and planetary materials to improve our understanding of the generation of planetary-scale isotopic anomalies for Ti and Sr. By comparing the results with data for other elements, we develop a quantitative model for the origin of planetary-scale anomalies and the causes for the offset between the non-carbonaceous (NC) and carbonaceous (CC) nebular reservoirs in the context of the evolution of the early solar system.

## 2. Samples and Methods

The leachate samples investigated here for their multi-elemental concentration and Ti and Sr isotopic composition are aliquots of a sequential dissolution of ~16.5 g whole-rock powder of the carbonaceous chondrite Murchison, which was performed at the University of Chicago using the following sequence:

L1:     9 M HAc (acetic acid), 1 day, 20 °C;
L2:     4.7 M $HNO_3$, 5 days, 20 °C;
L3:     5.5 M HCl, 1 day, 75 °C;
L4:     13 M HF – 3 M HCl, 1 day, 75 °C;
L5:     13 M HF – 6 M HCl, 3 days, 150 °C;
L6:     Insoluble residue.

The aliquot of the insoluble residue (L6) was fused by a $CO_2$ laser under a reducing atmosphere before digestion in $HNO_3$–HF–$HClO_4$ (see Burkhardt et al. 2012b). Molybdenum and W isotopic compositions on the same aliquots have been reported in Burkhardt et al. (2012a, b), and Ca, Cr, Ru, and Os isotope compositions obtained on different aliquots of the sequential digestion are given in Chen et al. (2011), Papanastassiou et al. (2010), Fischer-Gödde et al. (2014), and Reisberg et al. (2009), respectively.

The five Allende CAIs analyzed here for their multi-elemental concentration and Ti and Sr isotopic composition have also been previously investigated for Mo and W isotope systematics (Burkhardt et al. 2008; 2011), and for two of them (A-ZH-2, A-ZH-4) Sr isotope data was reported in Hans et al. (2013). The CAIs A-ZH-1 to A-ZH-4 are typical large coarse-grained type B igneous inclusions consisting mainly of melilite and fassaite, while CAI A-ZH-5 has a more fine-grained appearance with voids and alteration features (Burkhardt et al. 2008).

After digestion and treatment with aqua regia, all samples were dried down and redissolved in 6M HCl–0.01M HF. At this point ~0.4-5% aliquots were taken for concentration measurements. The remaining solutions were processed through ion exchange chromatography to separate Sr, Ti (this study), and Mo and W (Burkhardt et al. 2008; 2012a; b).

Strontium, along with the main matrix elements, was separated from the high field strength elements, Mo, and W on a primary cation exchange column (14 mL AG50W-X8 200-400 mesh) with 1M HCl–0.1M HF (Patchett and Tatsumoto 1980). Strontium and the matrix elements were washed off this column with 6M HCl. After drying down, Sr was separated from the matrix and Rb in 2.5M HCl on an ion exchange column filled with 3.6 mL BioRad AG50W-X8 (200-400 mesh). After evaporation, the Sr cuts were re-dissolved in 3M $HNO_3$ and further purified using a 200 µL Eichrom Sr resin column (Hans et al. 2013). Rubidium was washed off this column in 3M $HNO_3$ together with the matrix elements, while Sr was eluted with 0.05M $HNO_3$. After conversion into chloride form the Sr was then ready for measurement by TIMS. The total procedural blank for the Sr separation was ~14 pg, which is negligible given the high Sr contents of the samples.

The elution cut containing HFSE, Mo, and W was dried down and re-dissolved in 1M HF (leachates) or 1M HCl–0.1M HF (CAIs) and loaded on columns filled with AG1-X8 (100-200 mesh) anion exchange resin. For the leachate samples the remaining matrix was washed off these columns in 1M HF, followed by elution of Ti (+Zr, Hf) in 6M HCl–0.06M HF, before W was eluted in 6M HCl–1M HF, and finally Mo in 3M $HNO_3$. The matrix of the CAI samples was washed off in 0.5M HCl–0.5M HF, Ti (+Zr, Hf) in 3.6M HAc–2% $H_2O_2$, W in 6M HCl–1M HF, and Mo in 6M $HNO_3$–0.2M HF, respectively (Burkhardt et al. 2008; 2012b). Titanium was further purified in Chicago following the methods laid out by Zhang et al. (2011). After drying down, the cuts were taken up in aqua regia, dried and re-dissolved in 12M $HNO_3$ (+ traces of $H_3BO_3$) and loaded on columns prepacked with 2 mL Eichrom DGA resin. After rinsing off the matrix in 12M $HNO_3$, Ti was eluted with a mixture of 12M $HNO_3$– 1wt.% $H_2O_2$ while Zr and Hf stayed on the column. Titanium cuts were dried with concentrated HF and finally loaded on pre-cleaned anion exchange columns (AG1-X8, 0.8 mL, 200-400 mesh) in 4M HF. After washing off the matrix and V, Ti was eluted with 9M HCl–0.01M HF, and, after conversion to 0.3M $HNO_3$–0.0014M HF solutions, was ready for measurement by MC-ICPMS.

The Ti isotope measurements were performed using the ThermoFisher Scientific Neptune MC-ICP-MS at the Origins Lab in Chicago in high resolution mode (Zhang et al. 2011). Solutions containing about 275 ppb Ti were introduced through a Cetac Aridus II desolvating system, resulting in a ~$4.5 \times 10^{-10}$ A ion beam on $^{48}$Ti. Measurements consisted of a two-line data acquisition scheme. The first line measured all Ti isotopes as well as a Ca interference monitor on mass 44, and the second line measured masses 49 and 50, as well as V and Cr interference monitors on masses 51 and 53. Measurements were performed on the peak shoulder to resolve polyatomic isobaric interferences of $^{22}Ne^{22}Ne^+$ on mass 44, $^{36}Ar^{14}N^+$ on mass 50 and $^{40}Ar^{13}C^+$ on mass 53. Each measurement consisted of a 30 s baseline (deflected beam) followed by 20 isotope ratio measurements of 16 s each. Interference corrections were small (maximum Ca correction was -0.47 $\varepsilon^{48}$Ti) and well within the previously defined correctable range (Zhang et al. 2011), or negligible (V, Cr, *i.e.* <0.05 $\varepsilon^{50}$Ti). The Ti isotope results are given as $\varepsilon^i$Ti relative to bracketing measurements of the Origins Lab OL-Ti standard ($\varepsilon^i$Ti = $[(^iTi/^{47}Ti)_{sample}/(^iTi/^{47}Ti)_{OL-Ti} -1] \times 10^4$; with i = 46, 48, 50). Instrumental mass bias was corrected using $^{49}Ti/^{47}Ti$ = 0.749766 as normalizing ratio with the exponential law. The external reproducibility (2 s.d.) of the standard during the analytical session was ±0.22 $\varepsilon^{46}$Ti, ±0.12 $\varepsilon^{48}$Ti, and ±0.38 $\varepsilon^{50}$Ti. The accuracy and reproducibility of the Ti isotope measurements was assessed by analyses of the terrestrial rock standard BHVO-2 and the CV3 chondrite Allende, both of which agree with previously published results.

Strontium isotope measurements were performed using the ThermoFisher Scientific Triton TIMS at ETH Zurich. For each measurement about 500 ng of Sr was loaded on a previously out-gassed (4 A for 2 hours with a 30 min peak at 4.5 A) Re single filament.

Approximately 1 µL of sample in 6M HCl was mixed with 1 µL of TaCl$_5$ activator containing phosphoric acid solution prior to loading. Two fine strips of Parafilm were used to prevent spreading of the sample solution. The samples were then loaded with a Hamilton syringe and dried down at 0.6 A. The Parafilm strips were burnt off at 1.2 A and the filament then glowed briefly to dull red at 2 A. Rubidium interferences were monitored by measuring $^{85}$Rb. Some samples contained minor Rb, but this was burnt off the filament prior to the Sr isotopic measurements, such that Rb interference corrections were negligible for all samples. Depending on sample size, Sr isotopes were measured at $^{88}$Sr intensities between 5×10$^{-11}$A and 2×10$^{-10}$A for 8-10 h and consisted of 800 cycles. Baselines were measured for 30 seconds prior to each block of 20 cycles.

The $^{84}$Sr/$^{86}$Sr measurements were performed by either a one-line static acquisition scheme or a two-line acquisition scheme. In the latter case, $^{84}$Sr was collected in the cups L1 and L3, with all other Sr isotopes collected accordingly in each magnet setting. Integration times were 8 s for each line with a 3 s idle time between mass jumps. The $^{84}$Sr/$^{86}$Sr ratios acquired in each line were corrected for instrumental mass fractionation with the exponential law using the $^{86}$Sr/$^{88}$Sr ratio acquired in the same line and assuming $^{86}$Sr/$^{88}$Sr = 0.1194. A 'multi-static' $^{84}$Sr/$^{86}$Sr ratio can then be obtained by averaging the two static $^{84}$Sr/$^{86}$Sr ratios acquired in the two lines. A dynamic $^{84}$Sr/$^{86}$Sr ratio is obtained by correcting the $^{84}$Sr/$^{86}$Sr ratio measured in the first line for mass fractionation using the $^{86}$Sr/$^{88}$Sr obtained in the second line. With this latter procedure, potential biases in cup efficiencies and gain factors of the amplifiers effectively cancel out (Hans et al. 2013) and yield more reliable results compared with static measurements. Repeated measurement of the NBS 987 Sr standard over the course of this study yielded ratios of $^{84}$Sr/$^{86}$Sr = 0.0564943±0.0000016 for the dynamic measurements (2 s.d.) and $^{84}$Sr/$^{86}$Sr = 0.0564925±0.0000025 for the multi-static measurements (2 s.d.).

Concentration measurements on digestion aliquots were performed on a ThermoFisher Scientific Element II quadrupole ICP-MS at the Institute für Planetologie at the University of Münster for the leachate samples, and an Elan quadrupole ICP-MS at ETH Zürich for the CAIs. The samples were introduced into the plasma without prior chemical separation and concentrations were calculated by comparing the count rates of the samples to multi-element standard solutions. The uncertainties on the concentrations are estimated for each element individually by comparing the concentrations of BCR-2, BHVO-2, JA-2, DTS-2a and Allende standard solutions measured alongside the unknowns at similar or lower intensity relative to literature reference values (Jochum et al. 2015; Jarosewich et al. 1987, Stracke et al. 2012).

## 3. Results

### 3.1. Leachate concentration data

The concentrations of the Murchison leachates are given in Table 1 and are shown in Figures 2 and 3 for selected elements. The leachate concentrations were calculated by dividing the total quantity of the element in the leachates by the total mass of the Murchison sample (~16.5 g), such that the sum of the concentrations in the individual leachates should add up to the bulk Murchison concentration. Except for elements that have been partly lost by evaporation during sample preparation (*e.g.,* Ru), the agreement with average CM chondrite data from the literature is very good, *e.g.,* for Ti and Sr with 595±9 and 9.7±0.4 µg/g here *vs.* 580 and 10.1 µg/g in Wasson and Kallemeyn (1988), respectively. This suggests that our measurements are accurate and that all components of Murchison were digested in the leaching procedure. Leach step 1 (9M acetic acid at room temperature) contains 50% of the total mass of the elements analyzed. Nearly all of the highly soluble lithophile elements Na (95%), K (84%), Sr (85%), and the majority (>50%) of In, Ca, Mn, Fe, Zn, Rb, Cd, Ba, LREE, and U present in Murchison are dissolved in the acetic acid leachate. This acid attacked halites, carbonates, phosphates, soluble organic matter, as well as some metal and silicates. The CI chondrite normalized REE abundance reveals subtle enrichments of LREE vs. HREEs and a positive Eu anomaly in the acetic acid leachate (Fig. 3a). Leach step 2 (4.7M $HNO_3$ at room temperature) contains 37% of the total mass of the analyzed elements. Most abundant is Fe (32% of total Fe), followed by Mg (49%), Ga (86%), Ni (51%) and Al (54%). This leachate contains the majority of all Tl (97%), Cu (96%), Pb (89%), Ga (86%), Bi (80%), Mo (68%) Co (60%), Rh (54%), Ni (51%), and major fractions of V, Ag, Ru, Re, and Os, and thus indicates digestion of metals and sulfides along with some oxides and silicates. The REE abundances are about a factor of 3-5 times lower than in the acetic acid and no LREE enrichment or Eu anomaly is visible (Fig 3a). Leachate 3 (5.5M HCl at 75°C) contains 6.4% of the total mass of the analyzed elements. Dominant are Mg (9.3% of total Mg), Fe (1.3%), Ca (20%) and Ti (35%). The REE are more abundant than in leach step 2, but lower than in leach step 1. They display a deficit in LREE relative to HREE and a negative Eu anomaly. Leach step 4 (13M HF–3M HCl at 75°C) dissolved 1.9% of the total mass of the analyzed elements. Most abundant is Mg (3% of total Mg), followed by Ca (6.6%) and Fe (0.3%). Due to HF, the abundances of HFSEs (Ti 35%, Zr 23%, Nb 76%, Hf 32%, Ta 80%, W 48%) in this leachate are high, while the REEs are almost two orders of magnitude lower than in the leachate 3. Leachate 5 (13M HF–6M HCl at 150°C) contains the least amount of material and dissolved only about 0.34% of the total mass of the elements analyzed. Only V, Cr, Zr, Nb, Hf and Ta are present in significant concentrations (>5% of total); while the REE abundances are down to ~0.1% of the chondritic values. The insoluble residue left after leach step 5 represents about 7.8% by weight of the Murchison starting powder and is strongly enriched in insoluble organic matter and highly refractory phases such as oxides and carbides. Fusion by laser under reducing conditions volatilized the organic (and other potentially remaining volatile) compounds and left an ultra-refractory residue of ~2.5% of the initial Murchison starting material. The residue is dominated by Mg (10% of total Mg) and Al (8%) and contains significant amounts of Pd (41%), Pt (38%), Zr (25%), Sc, Y, Re, Os, Ir, REE, and Hf (8-18%). Given the similar geo/cosmochemical behavior of Zr, Hf, and Ti the fraction of the latter in the refractory residue (0.3%) is somewhat lower than expected. This may indicate the formation of insoluble Ti-oxides during chemical processing of this sample, possibly through oxidation of $TiCl_4$ during dry-downs with $HClO_4$. The REE pattern of the residue shows a LREE depletion and a negative Eu anomaly, similar to the pattern of leachate 3. Integrating the REE concentrations of leachate 1 to 6 gives a flat REE pattern for the cumulative total with an enrichment of ~1.2 relative to CI (Fig. 3a), consistent with average bulk CM literature values (Wasson and Kallemeyn 1988).

*3.2. CAI concentration data*

The concentration data of the Allende CAIs are given in Table 2 and CI chondrite normalized REE patterns are shown in Figure 3b. As expected for refractory inclusions, all CAIs are strongly enriched in refractory elements, with enrichments ranging from 5×CI for U in A-ZH-2 to 25×CI for Sr in A-ZH-3. The average enrichment factors for CAIs A-ZH-1 to -5 relative to CI as based on Y, Zr, REE (except Eu), and Th concentrations are 15.0, 16.6±1.1, 13.9±0.6, 13.5±1.0, 15.5±1.6, respectively. The enrichment factors of Mo, W, and U are significantly lower than those of the other refractory elements in all CAIs. While for U this may relate to its relative volatility relative to other refractory elements, the relative depletions of Mo and W indicate volatility-controlled loss in oxidizing conditions (Fegley and Palme 1985). In addition, the 2+ elements Sr, Ba, and Eu (melilite compatible) are all depleted in CAI A-ZH-5. The REE patterns of the CAIs are flat, except for a positive Eu anomaly in the type B CAIs A-ZH-2 to -4 (group I pattern; Mason and Martin, 1977) and a negative Eu and Yb anomaly for A-ZH-5 (group III pattern). Europium and Yb are the most volatile REEs during condensation from a gas of solar composition (Boynton 1975; Davis and Grossman 1979). Furthermore, CAI A-ZH-5 is enriched in volatile Rb (1.7×CI), but not in Pb. The presence of alkalis is also consistent with extensive aqueous alteration previously documented in fine grained CAIs. Taken together, this indicates different nebular conditions during formation or processing of A-ZH-5 and the other CAIs in the circumsolar disk.

*3.3 Leachate isotope data*

The Ti and Sr isotopic data for the Murchison leachates are reported in Table 3 and are shown in Figures 4-6. The leachates exhibit significant mass-independent internal Ti and Sr isotopic heterogeneity, ranging from -5.10±0.23 $\varepsilon^{50}$Ti and +0.62±0.17 $\varepsilon^{84}$Sr in the first leachate to +10.54±0.26 $\varepsilon^{50}$Ti and -16.18±0.47 $\varepsilon^{84}$Sr in the residue, respectively. For a given sample the magnitude of the Ti anomaly decreases in the order $\varepsilon^{50}$Ti > $\varepsilon^{46}$Ti > $\varepsilon^{48}$Ti. The anomalies in $\varepsilon^{48}$Ti and $\varepsilon^{46}$Ti (Fig. 6a), and $\varepsilon^{48}$Ti and $\varepsilon^{50}$Ti (Fig. 6c) are negatively correlated, the ones in $\varepsilon^{50}$Ti and $\varepsilon^{46}$Ti (Fig. 6b) are positively correlated, however neither correlation defines a straight line in $\varepsilon^{i}$Ti and $\varepsilon^{j}$Ti mixing space. These results are qualitatively consistent with Ti isotope data obtained by the sequential leaching of the Orgueil CI chondrite (Trinquier et al. 2009). The weighted average of the leachates yields anomalies of +0.48±0.10 $\varepsilon^{46}$Ti, +0.01±0.08 $\varepsilon^{48}$Ti, +3.10±0.17 $\varepsilon^{50}$Ti, and +0.35±0.18 $\varepsilon^{84}$Sr (Fig. 4), all of which are fully consistent with literature data for bulk Murchison (Trinquier et al. 2009; Zhang et al. 2012; Moynier et al. 2012; Table 3). Contrary to the individual leachates, the weighted average plots on a linear array defined by bulk planetary materials, chondrules, and CAIs in $\varepsilon^{50}$Ti vs. $\varepsilon^{46}$Ti space (Fig. 6b).

The overall trend from slightly positive to strongly negative $\varepsilon^{84}$Sr anomalies observed here is qualitatively comparable to the results of other sequential digestion studies of Murchison (Qin et al. 2011; Yokoyama et al. 2015). The $^{87}$Sr/$^{86}$Sr of the leachates is dominated by decay from $^{87}$Rb ($t_{1/2}$=49.5 Ga) and ranges from unradiogenic 0.702120 in leachate 3 to radiogenic 0.753512 in leachate 2. For comparison, the initial $^{87}$Sr/$^{86}$Sr ratio at solar system birth is estimated to be 0.69875±0.000008 (Hans et al. 2013), indicating that even the least radiogenic leachate fraction contains radiogenic $^{87}$Sr from $^{87}$Rb-decay. The $^{87}$Sr/$^{86}$Sr ratio of the individual leachates are only poorly correlated with their measured $^{87}$Rb/$^{86}$Sr ratio, as indicated by the significant scatter around a 4.567 Ga chondrite reference line in a Rb-Sr isochron diagram (Fig. 5a). This behavior has been observed in other leachate studies as well (Qin et al. 2011; Yokoyama et al. 2015), and suggests incongruent dissolution of Rb and Sr during the leaching procedure (note that Murchison is an observed meteorite fall and Rb/Sr mobilization by terrestrial weathering is largely irrelevant). This is supported by the observation that the weighted average $^{87}$Rb/$^{87}$Sr and $^{87}$Sr/$^{86}$Sr of the leachates is in the range of bulk Murchison literature values and plots well on the 4.567 Ga isochron (Fig. 5a).

*3.4. CAI isotope data*

The Ti and Sr isotopic data for the Allende CAIs are provided in Table 4 and are shown in Figures 4-6. All CAIs exhibit significant and nearly uniform mass-independent non-radiogenic Ti and Sr isotopic anomalies, ranging from +9.12±0.13 $\varepsilon^{50}$Ti and +1.44±0.11 $\varepsilon^{84}$Sr in A-ZH-1 to +9.51±0.07 $\varepsilon^{50}$Ti and +1.07±0.74 $\varepsilon^{84}$Sr in A-ZH-5. No resolvable differences in Ti or Sr anomalies are present between the group I (A-ZH-1 to -4) and group III (A-ZH-5) CAIs. Anomalies in $\varepsilon^{46}$Ti, $\varepsilon^{48}$Ti, and $\varepsilon^{50}$Ti are positively correlated, with $\varepsilon^{46}$Ti and $\varepsilon^{48}$Ti values ~5.9 and ~19 times smaller than $\varepsilon^{50}$Ti. Compared to previous Ti isotope studies on Allende CAIs our data show less variability, but are fully consistent with a mean anomaly peak at ~+9 $\varepsilon^{50}$Ti (Niederer et al. 1985; Niemeyer and Lugmair 1984; Heydegger et al. 1982; Leya et al. 2009; Trinquier et al. 2009; Williams et al. 2016; Davis et al. 2018) (Figs. 4c, 6). In $\varepsilon^{50}$Ti vs. $\varepsilon^{46}$Ti space, CAIs, chondrules, and bulk planetary bodies plot on a common regression line (Fig. 6b), while in $\varepsilon^{48}$Ti vs. $\varepsilon^{46}$Ti and $\varepsilon^{48}$Ti vs. $\varepsilon^{50}$Ti space CAIs exhibit scatter beyond analytical uncertainty (Fig. 6a, c).

The $\varepsilon^{84}$Sr anomalies of the CAIs are indistinguishable within uncertainty and are well within the anomaly range reported in other studies (Moynier et al. 2012; Brennecka et al. 2013; Myojo et al. 2018). Variations in $^{87}$Sr/$^{86}$Sr range from 0.699478 in the coarse-grained Type B CAI A-ZH-2 to 0.718868 in the more fine-grained and volatile rich A-ZH-5, and correlate with Rb/Sr as expected for a ~4.567 Ga age (Fig. 5b).

## 4. Discussion

In the following sections we will attempt to identify the main drivers responsible for producing planetary-scale isotope variations. We investigate how planetary-scale anomalies relate to the mineral-scale isotopic heterogeneity inherited from the solar system's parental molecular cloud, and whether their formation can be accounted for solely by processes within the accretion disk itself or instead require a change in the isotopic composition of the infalling cloud material. We start by concentrating on Ti, and provide a global assessment of the relations of Ti isotopic anomalies among presolar grains, hibonites, CAIs, leachates, chondrules, matrix, and bulk planetary bodies (section 4.1). The same is then done for Sr (Section 4.2), before we expand our analysis to multi-elemental isotope space (Section 4.3), discuss the origin of the isotope anomalies with respect to disk processing and cloud heterogeneity (section 4.4), and finally synthesize our inferences in a global early solar system evolution model (Section 4.5).

*4.1. Titanium isotope anomalies in meteoritic materials*

Owing to the early discovery of Ti isotope anomalies in CAIs (Heydegger et al. 1979) and the relatively high abundance of Ti in most meteorites and chondrite components, Ti is among the elements for which isotopic anomalies have been most extensively characterized in planetary materials. However, since multiple processes and sources contribute to the Ti isotopic inventory, interpreting the meteoritic record is not straight-forward (*e.g.*, Steele and Boehnke 2015). Measurements on hundreds of presolar grains reveal percent-level Ti isotope variations among the grains (*e.g.*, Gyngard et al. 2018; Nittler et al. 2008). The Ti isotope variability in mainstream SiC grains (the most abundant type of presolar grains, which are thought to derive from AGB stars and are characterized by large *s*-process enrichments) mainly reflect the effect of galactic chemical evolution, resulting in the increase of the secondary nuclides $^{46}$Ti, $^{47}$Ti, $^{49}$Ti, and $^{50}$Ti relative to the early-formed primary nuclide $^{48}$Ti through time (Alexander and Nittler 1999), and, more subdued, slow *n*-capture processes in the parental AGB stars (Wasserburg et al. 2015).

Applying the internal $^{49}$Ti/$^{47}$Ti normalization scheme used here for the leachates and CAIs to the presolar SiC grain data results in mean apparent Ti anomalies of +704±47 (2s.e.) $\varepsilon^{46}$Ti, -396±34 $\varepsilon^{48}$Ti, and +503±84 $\varepsilon^{50}$Ti for mainstream SiC, and +2071±1228 $\varepsilon^{46}$Ti, -1674±535 $\varepsilon^{48}$Ti, and -2688±1123 $\varepsilon^{50}$Ti for SiC X-grains (rare, most likely of supernova origin), respectively (Fig 6). Mixing lines between terrestrial Ti and mainstream SiC have a negative slope in $\varepsilon^{48}$Ti vs. $\varepsilon^{46}$Ti and $\varepsilon^{48}$Ti vs. $\varepsilon^{50}$Ti space, and a positive slope for $\varepsilon^{50}$Ti vs. $\varepsilon^{46}$Ti (Fig. 6). In contrast to the SiC grains, hibonite-rich CAIs most commonly encountered in CM chondrites are thought to have mostly formed by condensation/evaporation processes in the nascent solar nebula. Hibonites exhibit more subdued anomalies in $\varepsilon^{46}$Ti and $\varepsilon^{48}$Ti, but cover a range of ~3000 in $\varepsilon^{50}$Ti (*e.g.*, Ireland et al. 1988; Kööp et al. 2016). Titanium anomalies in FUN CAIs are uncorrelated and restricted to a range of <100 $\varepsilon$ (Niederer et al. 1985; Kööp et al. 2018). "Normal" CAIs from CV, CO, CR, CK, and ordinary chondrites span a range of +2 to +15 $\varepsilon^{50}$Ti, with a prominent peak at about +9 $\varepsilon^{50}$Ti (Trinquier et al. 2009; Leya et al. 2009; Williams et al. 2016; Davis et al. 2018; Ebert et al. 2018; Render et al. 2019). Furthermore, while significant scatter beyond analytical precision is seen in CAIs in $\varepsilon^{48}$Ti vs. $\varepsilon^{46}$Ti, the anomalies in $\varepsilon^{46}$Ti vs. $\varepsilon^{50}$Ti in CAIs are well correlated ($R^2$=0.74) (Fig. 6).

The Ti isotope anomalies in the Murchison leachates are only marginally larger than the range covered by bulk bodies and CAIs. This indicates that in contrast to many other elements (see below) the leaching procedure applied here did not efficiently separate highly anomalous carriers relative to more normal Ti (Fig. 6). Nevertheless, the leachate data provide important information about the contributors to the intrinsic Ti isotopic heterogeneity of the Murchison chondrite. In an $\varepsilon^{46}$Ti–$\varepsilon^{48}$Ti diagram, the leachate anomalies are negatively correlated, whereas in $\varepsilon^{50}$Ti vs. $\varepsilon^{46}$Ti space they are positively correlated; however, most of the leachates do not plot on the linear array defined by bulk planetary materials and CAIs. In $\varepsilon^{i}$Ti vs. $\varepsilon^{j}$Ti space, binary mixing follows a straight line, and so the Ti isotopic compositions of the different leachates reveals multiple contributors to the intrinsic Ti isotopic heterogeneity of the Murchison chondrite. The variations in $\varepsilon^{46}$Ti and $\varepsilon^{48}$Ti can be largely accounted for by progressive enrichment of mainstream SiC grains from L1 to L6 (Fig. 6a). In particular the Ti isotopic composition of the residue (L6) is clearly shifted towards mainstream SiC. This fraction is enriched in SiC grains, as these are chemically resistant to the acids used in leaching steps 1-5. All heavy elements with isotopes produced by the *p*-, *s*-, and *r*-processes (Mo, Ru, Ba, Sm) measured in L6 or equivalent, chemically resistant residues show *s*-process enrichments (Fig. 11d of Dauphas and Schauble 2016), consistent with a large mainstream SiC contribution.

We note that CAIs with low $\varepsilon^{46}$Ti and $\varepsilon^{50}$Ti tend to have negative $\varepsilon^{48}$Ti anomalies, and thus some of the scatter in $\varepsilon^{48}$Ti in CAIs might be related to admixing of Ti from mainstream SiC (Fig 6a, c). Accounting for potential mainstream SiC contribution in L6 leaves a large spread in $\varepsilon^{50}$Ti in the leachates, indicating that the main $\varepsilon^{50}$Ti variation is caused by one or multiple carriers with large $\varepsilon^{50}$Ti and subdued $\varepsilon^{46}$Ti and $\varepsilon^{48}$Ti variability. Hibonites are an obvious candidate with these features, but given their sub-solar Ca/Ti ratio, they cannot account for the 1:1 correlation between $\varepsilon^{48}$Ca and $\varepsilon^{50}$Ti observed for bulk meteorites. Presolar perovskite (Dauphas et al. 2014) or titanite ($CaTiSiO_5$) are potential alternative carriers.

In summary, and consistent with previous conclusions (Niederer et al. 1980; Trinquier et al. 2009; Steele and Boehnke 2015; Davis et al. 2018), our leachate data reveal the contribution of at least three components with anomalous Ti isotopic compositions to the solar system's Ti isotopic inventory. The disparate leachate anomalies contrast with the linear correlation of bulk planetary bodies and CAIs in $\varepsilon^{46}$Ti - $\varepsilon^{50}$Ti space, indicating that the carriers responsible for the leachate anomalies most likely played only a minor, if any, role in the generation of the bulk meteorite and CAI anomalies. Since the Ti isotope anomalies in carbonaceous chondrites scale with the abundance of CAIs, the range of Ti anomalies in the CC reservoir is substantially reduced if CAIs are subtracted (Niemeyer, 1988; Trinquier et al. 2009; Leya et al. 2009; Burkhardt et al. 2016). For example, subtracting visible CAIs by mass-balance reduces the $\varepsilon^{50}$Ti anomaly of CV chondrites from ~+3.5 to ~+2.2 $\varepsilon^{50}$Ti (Burkhardt et al, 2016). This is similar to the anomaly of +1.85 $\varepsilon^{50}$Ti measured for CI chondrites, which contain no visible CAIs. The Ti isotopic variability among CC chondrites is, therefore, mainly controlled by the abundance of (still visible) CAIs in them. However, although the correction for the CAIs reduces the range of anomalies in the CC reservoir, a significant offset between the NC ($\varepsilon^{50}$Ti <0) and CC ($\varepsilon^{50}$Ti >1.85) reservoirs remains. Thus, this offset cannot be accounted for by the admixture of CAIs *sensu stricto*.

Chondrules from carbonaceous, ordinary, and enstatite chondrites (Gerber et al. 2017), together with chondrite matrix, bulk meteorites, and CAIs (Trinquier et al. 2009; Zhang et al. 2011) plot along a single $\varepsilon^{46}$Ti-$\varepsilon^{50}$Ti correlation line (Fig. 6b). Furthermore, chondrules from carbonaceous chondrites span a range between about -2 and +7 $\varepsilon^{50}$Ti, whereas EC and OC chondrules exhibit only limited scatter around their bulk meteorite Ti isotope compositions. The Ti isotope anomalies in the CC chondrules are correlated with their Ti concentration and size, while no such relation is observed in OC and EC chondrules. This was interpreted as evidence for the presence and heterogeneous distribution of CAI-like material among chondrule precursors in the formation region of carbonaceous chondrites (Gerber et al. 2017). These observations indicate that bulk carbonaceous chondrites do not only contain CAIs *sensu stricto*, but also reprocessed CAI-like material. Together, these two components exert a strong control on bulk planetary Ti isotope anomalies, and variable admixing of CAI-like refractory material to an average inner solar nebula component (as defined here by ECs and OCs) can account for a large part of the planetary-scale anomalies and the difference between non-carbonaceous (NC) and carbonaceous (CC) nebular reservoirs.

However, precisely quantifying the overall effect of CAI addition on Ti isotopic signatures in bulk carbonaceous chondrites is difficult, because the total amount of CAI-like material in individual chondrites is the sum of the visible CAIs plus the CAI-like material that was incorporated and reprocessed in chondrules and matrix. Moreover, the refractory dust involved may have had a chemical composition different from those measured in large CAIs. However, carbonaceous chondrites, including CI chondrites, are characterized by an excess of refractory lithophile elements relative to ECs and OCs (Grossman and Larimer 1974; Wasson and Kallemeyn 1988). Taking the Ti/Si ratio as a proxy for the excess refractory material in the CCs, we find that the $\varepsilon^{50}$Ti and Ti/Si of the bulk CCs are correlated, and plot on mixing lines between average ECs and OCs, and CAIs. The range of bulk scale Ti isotopic anomalies and Ti/Si ratios among carbonaceous chondrites can, therefore, entirely be accounted for by admixture of ~3-7% CAI-like material to an EC- or OC-like starting composition (Fig. 7a).

*4.2. Strontium isotope anomalies in meteoritic materials*

The four stable Sr isotopes have different contributions from the *p*-, *r*-, main *s*- and weak *s*-processes: $^{84}$Sr is exclusively produced in the *p*-process; $^{86}$Sr and $^{87}$Sr are pure *s*-process nuclides (radiogenic $^{87}$Rb with $t_{1/2}$~48 Gyr can be considered stable on the timescale of nucleosynthesis in stars during the lifetime of the Galaxy); and $^{88}$Sr is produced by the *s*- and *r*-process (Arlandini et al. 1999). Most of the *s*-process Sr derives from AGB stars, but a fraction is also synthesized by the weak *s*-process in massive stars during the pre-supernovae stage (Woosley and Heger 2007). The nature of the *r*-process in the Sr mass region is not well constrained, but when *r*-process is used in the following, it refers to the *r*-residual after subtraction of the main *s*-process from the average solar system composition.

Strontium isotope data for SiC grains extracted from Murchison matrix reveal percent-level deficits in $^{84}$Sr/$^{86}$Sr, resulting from an enrichment in *s*-process $^{86}$Sr relative to *p*-process $^{84}$Sr in these presolar grains (Nicolussi et al. 1998; Podosek et al. 2004). The first evidence for mass-independent Sr isotope anomalies in material processed within the solar system was reported for FUN CAIs, which show deficits of ~–8 $\varepsilon^{84}$Sr and excesses of ~+32 $\varepsilon^{84}$Sr (Papanastassiou and Wasserburg, 1978) after normalization to a fixed $^{88}$Sr/$^{86}$Sr ratio. "Normal" CAIs exhibit anomalies of about +1.3 in $\varepsilon^{84}$Sr (Brennecka et al. 2013; Hans et al. 2013; Moynier et al. 2012; Myojo et al. 2018; Paton et al. 2013; Charlier et al. 2017). Nucleosynthetic Sr isotope anomalies at the bulk meteorite scale seem to be restricted to CC bodies (Moynier et al. 2012; Paton et al. 2013; Hans et al. 2013; Yokoyama et al. 2015). At the current level of analytical precision, the Earth, Moon, Mars, eucrites, angrites, ECs, and OCs have indistinguishable relative abundances of non-radiogenic Sr isotopes, while carbonaceous chondrites exhibit slightly positive anomalies of up to +0.65 $\varepsilon^{84}$Sr (Fig. 7b). Note that for internally normalized Sr isotope data it is impossible to distinguish *p*-, *s*-, and *r*-process anomalies, because variations in $^{87}$Sr are dominated by $^{87}$Rb-decay, such that only three isotopes are left, two of which are used for internal normalization. Thus, the apparent variations in $\varepsilon^{84}$Sr seen in leachates, CAIs, and bulk bodies may be due to variations in either $^{84}$Sr, $^{86}$Sr, and $^{88}$Sr, or a combination thereof. Double spike or sample-standard bracketing measurements are of little help because the isotopic anomalies in CAIs and bulk meteorites are small with regard to mass-dependent variations arising from evaporation/condensation processes or fluid alteration. Nevertheless, several lines of evidence suggest that the -15 $\varepsilon^{84}$Sr anomaly of the insoluble residue (L6) reflects an excess in *s*-process isotopes. The insoluble residue of Murchison is enriched in presolar SiC produced in low mass AGB stars, and SiC grains reveal clear *s*-process signatures, including low $^{84}Sr/^{86}Sr$ ratios (Nicolussi et al. 1998; Podosek et al. 2004). Furthermore, isotopic analyses of residues remaining after dissolution of primitive chondrites reveal *s*-process enrichments for several elements, including Zr, Mo, Ba, W, and Os (Schönbächler et al. 2005; Burkhardt et al. 2012a; Qin et al. 2011; Burkhardt and Schönbächler 2015; Elfers et al. 2018). Based on a comprehensive study of nucleosynthetic isotope anomalies in Murchison leachates, Qin et al. (2011) also concluded that the large $\varepsilon^{84}Sr$ deficits in the insoluble residue reflected an excess in *s*-process Sr. Likewise, leachate studies on the SiC-rich carbonaceous chondrites Ivuna (Paton et al. 2013), Murchison, and Tagish Lake (Yokoyama et al. 2015) obtained qualitatively similar results to those found here, while leachates of the slightly metamorphosed and SiC-poor Allende (Yokoyama et al. 2015) did not show significant $\varepsilon^{84}Sr$ variations.

Figure 8 shows the $\varepsilon^{84}Sr$ values of the different leach steps in comparison to the $\varepsilon^{94}Mo$ values measured on the same aliquots. The $\varepsilon^{84}Sr$ and $\varepsilon^{94}Mo$ leachate anomalies are positively correlated, and roughly follow the expected co-variation resulting from a variable distribution of *s*-process nuclides, calculated using the formalism of Dauphas et al. (2004) and the *s*-process abundances of Bisterzo et al. (2014) and Arlandini et al. (1999). Note, however, that the *s*-process yields in these publications are relevant only for the nucleosynthesis in AGB stars, but up to 20-40% of $^{87}Sr$ and $^{86}Sr$ is produced by the weak *s*-process (Woosley and Heger 2007), while only 5% of $^{88}Sr$ was produced by this process (Travaglio et al. 2004). The acetic acid leachate (L1) plots significantly off the calculated $\varepsilon^{84}Sr$–$\varepsilon^{92}Mo$ *s*-process mixing line and has a much lower $\varepsilon^{84}Sr$ than expected for its $\varepsilon^{94}Mo$ (Fig. 8). This is likely the result of aqueous alteration on the parent body, whereby a large fraction of Sr from Murchison was mobilized and redistributed into carbonates (Riciputi et al. 1994) while the carrier(s) of Mo were less affected. In summary, the leachates provide evidence for intrinsic *s*-process Sr variability in Murchison, particularly in the later steps, whereas the early leach steps are dominated by Sr that has been homogenized by fluid mobilization on the parent body.

All CAIs investigated here are characterized by a positive $\varepsilon^{84}Sr$ anomaly, consistent with previously reported results for CAIs from CV3 (Moynier et al. 2012; Hans et al. 2013; Brennecka et al. 2013; Paton et al. 2013; Myojo et al. 2018) and CK3 (Shollenberger et al. 2018) chondrites. Thus, elevated $\varepsilon^{84}Sr$ seem to be a common feature of "normal" (*i.e.*, non-FUN) CAIs. As outlined above, positive $\varepsilon^{84}Sr$ may reflect either a *p*-excess, *r*-excess, or *s*-deficit (Papanastassiou and Wasserburg, 1978). Paton et al. (2013) interpreted the elevated $^{84}Sr/^{86}Sr$ of CAIs to be due to a deficit in *s*-process Sr, which these authors inferred to reflect the formation of CAIs from essentially SiC-free material. In contrast, other studies argued that the positive $\varepsilon^{84}Sr$ of CAIs is due to an excess in *r*-process Sr, since the same CAIs analyzed for Sr have been found to exhibit an *r*-process excess in the neighboring element Mo (Burkhardt et al. 2011; Brennecka et al. 2013). However, for Ru, which is the next neighboring element, the isotope anomalies in CAIs seem more consistent with *s*-process rather than *r*-process variations (Chen et al. 2010). CAI A-ZH-5 is particularly interesting in this respect, because although its $\varepsilon^{84}Sr$ (and $\varepsilon^{50}Ti$) anomaly is within uncertainty identical to other "normal" CAIs, its Mo isotopic signature differs significantly (Burkhardt et al. 2011). In addition to an *r*-excess similar to that of other CAIs, A-ZH-5 exhibits a large *s*-process Mo deficit, which, when translated into a deficit in *s*-process Sr, would correspond to an $\varepsilon^{84}Sr$ of ~+7.5, much higher than the observed value of ~+1.1 (Table 4). Thus, at least in terms of *s*-process variability, Mo and Sr isotope anomalies in CAIs can be decoupled. As such, the data for A-ZH-5 indicate that whereas the homogeneous enrichment in *r*-process material relative to the terrestrial composition is a common feature of the CAI-forming reservoir, the *s*-process variability in Mo is a secondary trait. These observations are fully consistent with inferences from Mo isotope anomalies at the bulk meteorite scale, which show that *s*-process Mo variability exists within the NC and CC reservoirs, whereas the offset between them is caused by an approximately homogeneous *r*-process excess in the CC over the NC reservoir (Budde et al. 2016; Kruijer et al. 2017; Poole et al. 2017; Worsham et al. 2017).

The $\varepsilon^{84}$Sr anomalies in CC bodies range from ~+0.4 in CI to ~+0.65 in CV chondrites, whereas terrestrial rocks, ECs, OCs, angrites, HEDs, as well as lunar and martian samples have indistinguishable non-radiogenic Sr isotopic compositions at the current level of analytical precision (Moynier et al. 2012; Hans et al. 2013; Paton et al. 2013; Yokoyama et al. 2015). As for Ti, the Sr isotopic composition of the carbonaceous chondrites and the offset between the NC and the CC reservoir is mainly controlled by CAIs (and reprocessed CAI-like matter). The range of bulk-scale Sr isotopic anomalies and Si/Sr ratios of CC bodies and their offset from the NC reservoir can be accounted for by admixture of ~4–10% of CAI-like materials to an EC- or OC-like starting composition (Fig. 7b).

*4.3. Origin and relation of isotope anomalies in meteoritic components and bulk meteorites in multi-elemental parameter space*

Due to continuous analytical progress it is now evident that chondrites display intrinsic nucleosynthetic isotopic variability in virtually every element investigated. The sequential digestion of carbonaceous chondrites revealed anomalies in Ca, Ti, Cr, Sr, Mo, Ru, Pd, Ba, Nd, Sm, Er, Yb, Hf, W, and Os isotopes, and only a handful of elements such as Cd, Sn, and Te evaded the clear detection of internal isotope anomalies thus far. For elements beyond the Fe-peak, the individual leachate data are all well explained by excesses and deficits in *s*-process isotopes (see Fig. 12 in Dauphas and Schauble 2016), highlighting that *s*-process carriers play an important role in explaining the intrinsic chondrite/nebular variability. When the data are shown in multi-elemental isotope space it becomes evident that different elements behave differently during circumstellar condensation of their presolar carriers, nebular and parent body processing, and leaching. This leads to variable element ratios in the carriers and the leachates and explains why the leachate anomalies are well correlated for some elements, but not for others (Burkhardt et al. 2012b). This can be seen in Figure 9 in diagrams of $\varepsilon^{50}$Ti *vs*. $\varepsilon^{48}$Ca, $\varepsilon^{54}$Cr, $\varepsilon^{84}$Sr, $\varepsilon^{96}$Zr, $\varepsilon^{94}$Mo, $\varepsilon^{100}$Ru, $\varepsilon^{135}$Ba, $\varepsilon^{145}$Nd, $\varepsilon^{144}$Sm, $\varepsilon^{180}$Hf, $\varepsilon^{183}$W, and $\varepsilon^{186}$Os isotope anomalies in leachates and acid residues of the Murchison meteorite (equivalent diagrams with $\varepsilon^{84}$Sr and $\varepsilon^{94}$Mo on the abscissa are provided in supplementary Figures S1 and S2). In these diagrams, the shape of two-component mixing curves depends on the element concentration ratio in the mixing end-members. Only when the element ratio in both end-members is the same, the mixing curve is a straight line, in all other cases it is a hyperbola (Langmuir et al. 1978). Titanium isotope anomalies of the leachates are well correlated with isotope anomalies for geochemically similar elements like Zr, Hf, Nd, as well as with isotope anomalies for Mo and W. The general trend is from a *s*-deficit in L1 to an *s*-excess in the residue, suggesting common *s*-process carriers and limited fractionation of these elements during the leaching. Very poor, if any, correlation exists in the leachates for Ti and Ca, Cr, Ru, and Os although some of these elements, such as Ca and Ti, show tight correlations at the bulk meteorite scale (Dauphas et al. 2014; Schiller et al. 2015). The reason for the decoupling of leachate and bulk anomalies reflects the fact that these elements exhibit different geochemical and cosmochemical behavior, so that mixing between different mineral fractions tapped by the various leaching steps follow complex topologies in mixing space. Of note, the weighted average anomaly of the leachates is well within the range of bulk CC bodies for all elements analyzed. Furthermore, for no element pair can the offset between NC and CC meteorites be clearly attributed to the correlations seen in the leachates, further strengthening the observation that the NC-CC offset is decoupled from the dominant intrinsic anomalies as recorded in the leachates.

Instead, for many elements the CC meteorites are offset from the NC meteorites towards the composition of average CV3 CAIs. This can be seen in Figure 10 (and supplementary Figs

S3 and S4), which is equivalent to Fig. 9 (and S1 and S2), but focused on the bulk and CAI anomalies. For the anomaly pairs $\varepsilon^{50}$Ti vs. $\varepsilon^{84}$Sr, $\varepsilon^{96}$Zr, $\varepsilon^{135}$Ba, and $\varepsilon^{180}$Hf the offset of NC and CC and the range within each reservoir is well-explained by admixing of CAIs to EC and OC compositions. For these elements the bulk anomalies are thus dominated by the presence of CAIs, and within group *s*-process variability is minor. For Mo, Ru, Nd significant nucleosynthetic *s*-process variability within the NC and/or the CC reservoirs exist (Dauphas et al. 2004; Budde et al. 2016; 2019; Fischer-Gödde et al. 2015; Carlson et al. 2007; Burkhardt et al. 2016), such that the location and range of the NC and CC meteorites in $\varepsilon^{50}$Ti vs. $\varepsilon^{94}$Mo, $\varepsilon^{100}$Ru, and $\varepsilon^{145}$Nd space is a function of both, *s*-process variability and variable addition of CAI-like material. This is consistent with what has been found based on Mo isotopes alone: the variability in NC and CC reservoirs is due to *s*-process heterogeneity, whereas the offset between the reservoirs is due to the homogeneous excess of *r*-process (*i.e.*, CAI-like) material in the CC reservoir (Budde et al. 2016; Worsham et al. 2017; Poole et al. 2017; Kruijer et al. 2017).

Importantly, the isotopic offset of CC bodies from the NC bodies towards average CV3 CAIs is not limited to refractory elements. For example, in $\varepsilon^{50}$Ti vs. $\varepsilon^{54}$Cr, and $\varepsilon^{62}$Ni space the CC bodies are offset from NC bodies towards the isotopic composition of CAIs for all three elements (Fig. 10). Since Cr and Ni are highly depleted in CAIs relative to bulk chondrites, the mixing hyperbolas of CAIs and EC and OC compositions have a strong curvature and do not encompass the CC bodies. CAIs *sensu stricto* are, therefore, not a viable component to explain the offset between NC and CC bodies for the main component elements Ni and Cr (Nanne et al. 2019). Nevertheless, the Ni and Cr isotopic compositions of the CC bodies are shifted toward the composition of CAIs. This suggests that CAIs are the refractory part of an isotopically distinct early nebular reservoir, and that the less refractory part of this reservoir with an approximately CAI-like isotopic signature also plays a role in setting planetary-scale isotope anomalies, including the offset between the NC and CC reservoirs. This is supported by the $\varepsilon^{50}$Ti vs. $\varepsilon^{54}$Cr correlation observed for NC bodies, which defines a straight line pointing toward the composition of CAIs, and could, as such, be explained by variable admixture of material characterized by a CAI-like isotopic composition, but with broadly solar chemical abundances.

In summary, for many elements the offset in nucleosynthetic anomalies between NC and CC bodies is largely decoupled from intrinsic *s*-process dominated variability, but is instead dominated by CAIs and isotopically CAI-like material. As can be seen in Fig. 10, the bulk meteorite anomalies within the NC and CC reservoirs can be along the same trajectory as the NC-CC offset, particular for elements that are strongly enriched in CAIs (*e.g.*, Ti, Sr), while for other elements (*e.g.*, Mo) the NC-CC offset and the within-reservoir variation follow different trends. Also, for some elements the main driver for within-reservoir variability is different for the NC and CC reservoirs, as is evident from disparate trajectories of NC and CC trends. Thus, planetary-scale nucleosynthetic isotope anomalies are the result of several, sometimes superimposed processes that acted on at least two initially isotopically distinct nebular reservoirs. In the following we discuss how these distinct nebular reservoirs might have formed, and propose a potential scenario that can account for the elemental, isotopic, and structural diversity seen in planetary materials.

*4.4. Origin of isotopically distinct nebular reservoirs: nebular processing or molecular cloud heterogeneity?*

The main result of our analysis is that, although multiple processes contributed to the formation of planetary-scale isotopic heterogeneities, the difference between NC and CC bodies requires variable mixing between two isotopically distinct nebular reservoirs with approximately solar/chondritic chemical compositions. One of these reservoirs is isotopically similar to CAIs (we call this hypothetical reservoir *IC* for Inclusion-like Chondritic component) and the other is more similar to the NC reservoir. As CAIs are the oldest dated objects that formed in the solar system, it is likely that this reservoir formed first, whereas the NC-like reservoir formed later. As to whether the isotopic difference between these two reservoirs reflects selective thermal processing of presolar carriers during infall or disk evolution, or is inherited from primordial variability in the solar system's parental molecular cloud is more difficult to assess. This is because we lack a complete inventory of the carriers of nucleosynthetic isotope anomalies that were initially present in the molecular cloud, making it difficult to even qualitatively assess how physico-chemical processing during infall or in the disk might have affected the isotopic composition of distinct disk reservoirs. Nevertheless, several observations suggest that the isotopic difference between the two early disk reservoirs is inherited from the solar system's parental molecular cloud, and does not result from thermal processing. First, the NC-CC offset is decoupled from the presolar carriers dominating the internal nucleosynthetic heterogeneity of meteorites, indicating that the NC-CC offset cannot easily be attributed to the heterogeneous distribution of presolar carriers that are still present in primitive meteorites. Second, the NC-CC offset is not limited to the signature of a single nucleosynthetic process, and so does not reflect processing of a specific presolar carrier. Instead, the CC reservoir is characterized by an enrichment in nuclides produced in neutron-rich stellar environments, which are hosted in distinct carriers. To account for this observation by thermal processing would therefore require preferential processing of a specific set of carriers, while other carriers (e.g., *s*-process carriers) would not have been affected. Thus, thermal processing could only account for the NC-CC offset under very specific assumptions. Third, thermal processing would have likely generated non-solar chemical compositions, but the generation of the NC-CC dichotomy requires mixing of two reservoirs with approximately solar concentration ratios. In our opinion, these observations are best explained if the NC-CC dichotomy is an inherited isotopic heterogeneity of the parental cloud that was preserved, in diluted form, during infall and subsequent mixing and processing within the disk.

The Sun most likely formed in a stellar cluster (Adams 2010; Levison et al. 2010), and the presence and abundance of short-lived nuclides (in particular $^{26}$Al) in early solar system materials requires the addition of freshly synthesized supernova or Wolf-Rayet star-material shortly before or during the collapse of the molecular cloud core (e.g., Gaidos et al. 2009; Dwarkadas et al. 2017). Given these constraints it is conceivable that in this dynamic environment the isotopic composition of the collapsing cloud material varied at the scale of the cloud core, and dynamical simulations, albeit of coarse resolution, suggest that initial cloud heterogeneities can survive in the forming disk (Visser et al. 2009). The lack of $^{26}$Al in hibonite grains with large Ti and Ca isotope anomalies has been related to such cloud heterogeneities (Kööp et al. 2016). Thus, although it cannot be excluded that processing during infall or disk evolution contributed to the observed isotopic heterogeneities, our preferred interpretation for the origin of the two isotopically distinct nebular reservoirs and, hence, the NC-CC isotopic offset, is a change in the isotopic composition of the infalling cloud material.

*4.5. A qualitative model for isotopic and elemental heterogeneities in the early solar system*

Taking recent infall and disk evolution models (Pignatale et al. 2018; Yang and Ciesla 2012), the proposed formation location of CAIs near the young Sun, and the early formation of CAIs as a guide, we infer that the molecular cloud parcel with the CAI-like isotopic composition represents early infalling material, whereas the molecular cloud parcel with an NC-like isotopic composition represents later infalling material. The change in infall and associated disk evolution is qualitatively sketched in Fig. 11 (a similar model was proposed in Nanne et al. 2019). Because early infall is limited to regions close to the star, and the forming disk is rapidly expanding through viscous spreading, variably processed cloud materials with CAI-like isotope composition (including CAIs *sensu strictu*, as well as less refractory dust) will be transported outwards, or condensed during the outward transport of hot nebular gas (Fig. 11a). This process can account for the overabundance of refractory phases in outer solar system materials (Yang and Ciesla, 2012). During further infall the centrifugal radius of infalling material is expanding, and the primary signature of the early infall will be diluted (Fig. 11b). The dilution is highest in the inner part of the disk, because throughout accretion the mass fraction of infalling material is always highest close to the star. Thus, the isotopic signature of the later infalling NC material will dominate the inner part of the disk. By contrast, the outer disk retains a higher fraction of matter from the earliest disk (*i.e.*, with a CAI-like isotopic composition), and, because of the outward expansion of the centrifugal radius also receives a higher fraction of unprocessed primitive cloud material. Subsequent transport, mixing, and processing of material in the disk reduced the initial isotopic difference between the two early disk reservoirs to the observed NC-CC offset, and led to the generation of secondary features like the *s*-process variability in the NC and CC reservoirs. A strength of this model is that it naturally explains many of the observables in the meteorite record: the excess of refractory elements and the higher amount of unprocessed material in the CC reservoir; the higher relative abundance of SiC X grains to mainstream SiC grains in CC relative to NC chondrites (Lin et al. 2002); the presence of isotopically NC-like material with CAI-like condensation features in NC chondrites (Ebert et al. 2018); and the distinct isotopic compositions of the NC and CC reservoirs.

In our model, the isotopic composition of CAIs (which puzzled cosmochemists for decades because it cannot be related to a specific nucleosynthetic process for all elements) simply reflects the average composition of the early infalling molecular cloud material. In this framework, the similar O isotopic composition of the Sun and CAIs (McKeegan et al. 2011) are easily understood, because both formed predominately from the early infalling material. Obviously, while our two-component mixing scenario can explain many features of the elemental and isotopic meteorite record, it is only a simplified model of the processes that acted at the beginning of the solar system. Most likely the average CAI-like isotopic composition itself is a mixture of molecular cloud parcels, and even earlier infalling material may have had a different composition (*e.g.*, $^{26}$Al-free hibonites and FUN CAIs likely represent this even earlier-infalling material). However, the importance of these materials for understanding the isotopic variations among bulk meteorites is marginal. Therefore, based on our inferences the following testable prediction can be made: *for elements that do not show large systematic isotope anomalies in normal CAIs there will be no systematic offset between NC and CC meteorites*.

## 5. Conclusions

By systematically investigating the isotopic record of planetary materials in multi-element space, we evaluate the main controlling factors for the formation of planetary-scale nucleosynthetic isotope anomalies, and how these anomalies may relate to an isotopic heterogeneity of the solar system's parental molecular cloud. We find that the intrinsic (i.e., within-meteorite heterogeneity) isotope anomalies as sampled by acid leachates of primitive chondrites played only a minor, if any, role in the generation of Ti isotope anomalies in bulk meteorites and CAIs. We show instead that the Ti and Sr elemental and isotopic compositions of carbonaceous chondrites, and the isotopic offset between the NC and CC reservoir for these two elements, is mainly controlled by CAIs and reprocessed CAI-like material.

By expanding our analysis to other elements, we find that although the nucleosynthetic heterogeneity within the NC and CC reservoirs may have had variable origins (including nebular processing of *s*-process material), the systematic isotopic offset of the CC reservoir from the NC reservoir towards the composition of CAIs is a global feature and is not limited to refractory elements. This implies that the CC reservoir formed by mixing of two disk reservoirs having chondritic/solar chemical, but distinct isotopic compositions. In our preferred model, the isotopic difference between these two early disk reservoirs is a primordial feature that has been inherited from the solar system's parental molecular cloud during infall, whereby the early-infalling material was characterized by an approximately CAI-like isotopic composition (labeled IC for Inclusion-like Chondritic component), whereas later-infalling material had an NC-like isotopic composition. Subsequent mixing and processing within the disk resulted in additional isotopic variations within both reservoirs, and reduced the initial isotopic difference between them to the observed NC-CC isotopic offset. Planetary-scale nucleosynthetic isotope anomalies are, therefore, related to both, changes in the isotopic composition of the infalling material—ultimately leading to the isotopic offset between the NC and CC reservoirs—*and* differential processing and transport in the evolving circum-solar disk, resulting in isotopic variations within both reservoirs.

**Acknowledgments**

We are grateful to the Field Museum for providing the Murchison sample, and thank A. Meibom, H. Palme, and an anonymous reviewer for editorial handling and constructive comments. Also, we thank L. Kööp for sharing hibonite data. This study was supported by SNSF grant (PBE2PZ-145946) to CB, and SNSF grant (200021-113625) to BB, and NASA grants 359NNX17AE86G (LARS), NNX17AE87G (Emerging Worlds), and 80NSSC17K0744 360 (Habitable Worlds) to ND.

**Author contributions**

Element concentration and Ti isotope data were obtained by CB, Sr isotope data by UH under BB supervision. ND prepared the Murchison leachates, and encouraged CB to make use of the acquired data. CB interpreted the data, compiled literature data, performed calculations, and wrote the manuscript. The conceptual disk evolution model developed during discussions between TK and CB. All authors commented on the manuscript.

Table 1: Element concentration data of Murchison leachates

| Analyte | unit | L1 | L2 | L3 | L4 | L5 | L6 | Total | s % |
|---|---|---|---|---|---|---|---|---|---|
| Li | ng/g | 320 | 383 | 103 | 4.3 | 1.3 | 13 | 824 | 5 |
| Na | µg/g | 3110 | 90.6 | 48.6 | 6.36 | 4.26 | 3.53 | 3263 | 3 |
| Mg | µg/g | 31874 | 53182 | 10217 | 3280 | 142 | 10822 | 109518 | 8 |
| Al | µg/g | 2554 | 4566 | 527 | 76.22 | 39.6 | 715.96 | 8479 | 3 |
| K | µg/g | 291 | 47.9 | 2.56 | 2.50 | 0.81 | 1.30 | 346 | 6 |
| Ca | µg/g | 6586 | 991 | 2259 | 747 | 55.0 | 612 | 11250 | 5 |
| Sc | µg/g | 1.9 | 2.9 | 1.6 | n.d. | n.d. | 1.2 | 7.7 | 40 |
| Ti | µg/g | 22.7 | 143 | 210 | 207 | 9.67 | 1.99 | 595 | 5 |
| V | µg/g | 10.46 | 26.78 | 10.94 | 6.63 | 3.53 | 5.38 | 63.7 | 11 |
| Cr | µg/g | 645 | 833 | 300 | 424 | 244 | 158 | 2604 | 10 |
| Mn | µg/g | 760 | 417 | 151 | 159 | 14.4 | 9.3 | 1511 | 8 |
| Fe | µg/g | 120232 | 60488 | 6843 | 607 | 144 | 227 | 188541 | 10 |
| Co | µg/g | 182 | 289 | 7.1 | 0.60 | 0.21 | 0.49 | 479 | 12 |
| Ni | µg/g | 4403 | 4763 | 117 | 6.4 | 2.6 | 28.9 | 9321 | 11 |
| Cu | µg/g | 0.22 | 90.9 | 2.61 | 0.47 | 0.13 | 0.32 | 94.6 | 12 |
| Zn | µg/g | 129 | 38.7 | 2.34 | 0.86 | 0.41 | 0.10 | 172 | 15 |
| Ga | ng/g | 624 | 5794 | 199 | 63 | 77 | 1.2 | 6759 | 25 |
| Rb | ng/g | 885 | 653 | 20.0 | 1.82 | 0.66 | 0.88 | 1561 | 2 |
| Sr | ng/g | 8203 | 976 | 389 | 54.9 | 7.53 | 75.1 | 9705 | 6 |
| Y | ng/g | 722 | 229 | 548 | 7.86 | 3.30 | 288 | 1798 | 7 |
| Zr | ng/g | 797 | 836 | 253 | 1013 | 395 | 1108 | 4403 | 4 |
| Nb | ng/g | 1.3 | 2.1 | 20 | 279 | 46 | 20 | 369 | 30 |
| *Mo* | ng/g | *49.0* | *784.4* | *213.3* | *74.6* | *8.3* | *24.1* | *1154* | *0.1* |
| Ru | ng/g | 194 | 167 | 37.0 | 5.4 | 5.1 | 8.8 | 417 | 8 |
| Rh | ng/g | 30.7 | 70.2 | 15.9 | 1.99 | 0.11 | 11.1 | 130 | 10 |
| Pd | ng/g | 46 | 41 | 180 | 11 | n.d. | 190 | 468 | 30 |
| Ag | ng/g | 4.0 | 48.6 | 51.1 | 8.8 | 2.1 | 3.6 | 118 | 80 |
| Cd | ng/g | 306 | 103 | 7.46 | 3.90 | 1.21 | 1.71 | 423 | 80 |
| In | ng/g | 42.10 | 7.47 | 1.32 | 0.35 | 0.19 | 0.02 | 51.4 | 18 |
| Ba | ng/g | 908 | 631 | 37.3 | 2.75 | 0.75 | 19.9 | 1600 | 2 |
| La | ng/g | 206 | 42.7 | 53.9 | 0.91 | 0.35 | 17.0 | 321 | 5 |
| Ce | ng/g | 447 | 107 | 162 | 2.23 | 1.08 | 49.7 | 769 | 3 |
| Pr | ng/g | 62.3 | 15.2 | 27.8 | 0.33 | 0.17 | 8.6 | 114 | 3 |
| Nd | ng/g | 301 | 71.8 | 160 | 1.94 | 1.25 | 49.8 | 586 | 5 |
| Sm | ng/g | 95.1 | 23.8 | 55.5 | 1.00 | 0.46 | 21.1 | 197 | 6 |
| Eu | ng/g | 58.2 | 9.30 | 4.53 | 0.09 | 0.08 | 1.19 | 73.4 | 7 |
| Gd | ng/g | 114 | 34.3 | 81.2 | 0.94 | 0.69 | 34.3 | 266 | 16 |
| Tb | ng/g | 18.52 | 6.33 | 14.72 | 0.18 | 0.15 | 6.42 | 46.3 | 3 |
| Dy | ng/g | 129 | 42.9 | 100 | 1.57 | 1.06 | 47.0 | 321 | 4 |
| Ho | ng/g | 26.9 | 9.38 | 21.42 | 0.36 | 0.20 | 10.9 | 69.1 | 6 |
| Er | ng/g | 83.9 | 28.1 | 62.3 | 1.36 | 0.57 | 32.8 | 209 | 7 |
| Tm | ng/g | 12.9 | 3.95 | 8.83 | 0.12 | 0.06 | 4.77 | 30.7 | 6 |
| Yb | ng/g | 90.5 | 26.1 | 54.9 | 0.55 | 0.31 | 28.9 | 201 | 9 |
| Lu | ng/g | 12.3 | 4.2 | 8.1 | 0.17 | 0.04 | 5.35 | 30.2 | 5 |
| *Hf* | ng/g | *14.49* | *30.70* | *10.48* | *41.22* | *8.47* | *23.73* | *129* | *0.2* |
| Ta | ng/g | 0.14 | n.d. | 0.19 | 14.18 | 2.75 | 0.52 | 18 | 24 |
| *W* | ng/g | *7.83* | *20.85* | *27.42* | *59.43* | *3.00* | *4.50* | *123.0* | *0.8* |
| *Re* | ng/g | *16.0* | *22.0* | *2.3* | *0.50* | *0.30* | *6.0* | *46* | *15* |
| *Os* | ng/g | *117* | *265* | *69* | *12* | *20* | *63* | *547* | *3* |
| Ir | ng/g | 144 | 168 | 59 | 10.3 | 2.8 | 77 | 461 | 30 |
| *Pt* | ng/g | *93* | *243* | *200* | *21* | *5* | *339* | *900* | *1.9* |
| Tl | ng/g | 0.30 | 58.4 | 1.07 | 0.23 | 0.23 | 0.00 | 60 | 14 |
| Pb | ng/g | 97 | 1087 | 24 | 5.3 | 1.9 | 0.5 | 1216 | 50 |
| Bi | ng/g | 4.5 | 39 | 4.8 | 0.36 | 0.09 | 0.07 | 49 | 22 |
| Th | ng/g | 18 | 15 | 7.5 | 0.05 | 0.07 | 2.7 | 43 | 20 |
| U | ng/g | 5.8 | 1.3 | 1.6 | 0.07 | 0.10 | 0.60 | 9.5 | 28 |

Concentrations in the leachates were calculated by by dividing the quantity of the element in the leachate by the total mass of the Murchison sample. Uncertainty is estimated by average deviation of concentrations of BCR-2, BHVO-2, JA-2, DTS-2a, and Allende solutions (measured at similar or lower intensity along with the leachates) from literature reference values (Jochum et al. 2015; Jarosewich et al. 1987, Stracke et al. 2012). Values in italic were obtained by isotope dilution in Burkhardt et al. (2012a,b) (Mo, Hf, W) and Reisberg et al. (2009) (Re, Os, Pt).

Table 2 Element concentration data of Allende CAIs

| | unit | A-ZH-1 | A-ZH-2 | A-ZH-3 | A-ZH-4 | A-ZH-5 | s % | A-ZH-1 | A-ZH-2 | A-ZH-3 | A-ZH-4 | A-ZH-5 |
|---|---|---|---|---|---|---|---|---|---|---|---|---|
| | | | | | | | | | relative to CI | | | |
| Mass | mg | 120 | 180 | 70 | 200 | 90 | | | | | | |
| Type | | B | B | B | B | A | | | | | | |
| Group | | | I | I | I | III | | | | | | |
| Rb | µg/g | | *0.49* | 1.95 | *0.70* | 3.96 | 9 | | 0.21 | 0.84 | 0.30 | 1.71 |
| *Sr* | µg/g | | *186.20* | 194.87 | *137.10* | 35.04 | 15 | | 23.84 | 24.95 | 17.55 | 4.49 |
| Y | µg/g | | 26.15 | 23.31 | 22.00 | 23.44 | 14 | | 17.09 | 15.23 | 14.38 | 15.32 |
| Zr | µg/g | | 60.02 | 51.60 | 52.57 | 61.25 | 9 | | 16.58 | 14.25 | 14.52 | 16.92 |
| Nb | µg/g | | 2.35 | 4.23 | 2.68 | 5.73 | 8 | | 8.44 | 15.15 | 9.61 | 20.54 |
| *Mo* | µg/g | *8.27* | 7.89 | 6.59 | 6.75 | 5.78 | 0.1 | 8.50 | 8.11 | 6.77 | 6.94 | 5.94 |
| Ba | µg/g | | 59.47 | 30.68 | 42.37 | 7.36 | 11 | | 24.67 | 12.73 | 17.58 | 3.06 |
| La | µg/g | | 4.22 | 3.41 | 3.32 | 3.76 | 10 | | 17.43 | 14.09 | 13.70 | 15.55 |
| Ce | µg/g | | 11.13 | 8.75 | 8.32 | 9.68 | 11 | | 17.89 | 14.07 | 13.38 | 15.56 |
| Pr | µg/g | | 1.50 | 1.27 | 1.20 | 1.44 | 14 | | 15.87 | 13.40 | 12.67 | 15.21 |
| Nd | µg/g | | 7.39 | 6.28 | 5.88 | 7.24 | 15 | | 15.68 | 13.33 | 12.48 | 15.38 |
| Sm | µg/g | | 2.36 | 2.06 | 1.91 | 2.39 | 14 | | 15.53 | 13.55 | 12.59 | 15.76 |
| Eu | µg/g | | 1.12 | 1.18 | 0.94 | 0.31 | 11 | | 19.44 | 20.46 | 16.29 | 5.31 |
| Gd | µg/g | | 3.49 | 2.87 | 2.79 | 3.31 | 7 | | 17.05 | 14.02 | 13.62 | 16.16 |
| Tb | µg/g | | 0.61 | 0.50 | 0.48 | 0.58 | 9 | | 15.92 | 12.98 | 12.57 | 15.13 |
| Dy | µg/g | | 4.33 | 3.49 | 3.39 | 4.04 | 9 | | 16.98 | 13.70 | 13.30 | 15.84 |
| Ho | µg/g | | 0.92 | 0.76 | 0.74 | 0.88 | 9 | | 16.16 | 13.23 | 12.96 | 15.30 |
| Er | µg/g | | 2.73 | 2.25 | 2.23 | 2.59 | 8 | | 16.76 | 13.79 | 13.66 | 15.90 |
| Tm | µg/g | | 0.43 | 0.35 | 0.32 | 0.41 | 9 | | 16.33 | 13.57 | 12.42 | 15.69 |
| Yb | µg/g | | 2.58 | 2.44 | 2.18 | 1.80 | 9 | | 15.26 | 14.43 | 12.89 | 10.63 |
| Lu | µg/g | | 0.42 | 0.35 | 0.35 | 0.40 | 8 | | 16.59 | 13.73 | 13.93 | 15.90 |
| *Hf* | µg/g | *1.60* | 2.09 | 1.57 | 1.73 | 2.05 | 0.2 | 15.05 | 19.69 | 14.76 | 16.33 | 19.30 |
| Ta | µg/g | | 0.21 | 0.20 | 0.20 | 0.23 | 8 | | 14.68 | 14.13 | 13.89 | 16.14 |
| *W* | µg/g | *1.03* | 1.22 | 1.16 | 0.96 | 1.35 | 0.8 | 10.69 | 12.69 | 12.06 | 10.03 | 14.08 |
| Pb | µg/g | | 0.24 | 0.27 | 0.25 | 0.29 | 8 | | 0.09 | 0.10 | 0.10 | 0.11 |
| Th | µg/g | | 0.46 | 0.42 | 0.42 | 0.46 | 7 | | 14.82 | 13.56 | 13.52 | 14.69 |
| U | µg/g | | 0.05 | 0.07 | 0.05 | 0.10 | 7 | | 5.56 | 9.04 | 6.40 | 12.55 |

Measurements were performed on digestion aliquots containing between 0.4 to 1.4 % of the sample mass. Uncertainty is estimated by average deviation of concentrations obtained for BHVO-2 solutions (measured at similar or lower intensity as the CAIs) from literature reference values (Jochum et al. 2015). Values in italic were obtained by isotope dilution in previous studies (Burkhardt et al. 2008,2011; Hans et al. 2013). Normalization to CI chondrites was done using data of Lodders et al. (2009).

Table 3 Titanium and Sr isotope data of Murchison leachates

| Sample | Procedure | Ti [μg/g] ± 2σ | Rb [ng/g] ± 2σ | Sr [ng/g] ± 2σ | $^{87}Rb/^{86}Sr$ ± 2σ | $\varepsilon^{46}Ti^a$ ± 2σ | $\varepsilon^{48}Ti^a$ ± 2σ | $\varepsilon^{50}Ti^a$ ± 2σ | $\varepsilon^{84}Sr^b$ ± 2σ | $^{87}Sr/^{86}Sr$ ± 2σ |
|---|---|---|---|---|---|---|---|---|---|---|
| Murchison | | | | | | | | | | |
| L1 | 9M HAc 1day, 20°C | 22.7 ± 1 | 885 ± 18 | 8203 ± 492 | 0.31 ± 0.02 | -0.40 ± 0.11 | 0.30 ± 0.06 | -5.10 ± 0.23 | 0.62 ± 0.17 | 0.728748 ± 0.000002 |
| L2 | 4.7M HNO$_3$, 5 days, 20 °C | 143 ± 7 | 653 ± 13 | 976 ± 59 | 1.94 ± 0.12 | 0.18 ± 0.16 | 0.07 ± 0.08 | -1.04 ± 0.21 | 0.05 ± 0.19 | 0.753512 ± 0.000003 |
| L3 | 5.5M HCl, 1 day, 75 °C | 210 ± 11 | 20.0 ± 0.4 | 389 ± 23 | 0.15 ± 0.01 | 0.66 ± 0.08 | -0.07 ± 0.08 | 5.13 ± 0.17 | -1.05 ± 0.29 | 0.702120 ± 0.000004 |
| L4 | 13M HF 3M HCl, 1 day, | 207 ± 10 | 1.82 ± 0.04 | 54.9 ± 3 | 0.10 ± 0.01 | 0.55 ± 0.08 | 0.03 ± 0.08 | 4.46 ± 0.14 | -1.64 ± 0.67 | 0.703798 ± 0.000008 |
| L5 | 13M HF 6M HCl, 3 days, | 9.7 ± 0.5 | 0.66 ± 0.01 | 7.5 ± 1.0 | 0.25 ± 0.03 | 0.92 ± 0.14 | -0.12 ± 0.06 | 9.20 ± 0.18 | -2.70 ± 0.77 | 0.704667 ± 0.000008 |
| L6 | Insoluble residue, Laser f | 2.0 ± 0.1 | 0.88 ± 0.02 | 75.1 ± 5 | 0.034 ± 0.002 | 3.63 ± 0.10 | -1.92 ± 0.10 | 10.54 ± 0.26 | -16.18 ± 0.47 | 0.704959 ± 0.000006 |
| Total (conc.) or wt. ave. (ratios) | | 595 ± 9 | 1561 ± 15 | 9705 ± 423 | 0.47 ± 0.02 | 0.48 ± 0.10 | 0.01 ± 0.08 | 3.10 ± 0.17 | 0.35 ± 0.18 | 0.729826 ± 0.000002 |
| Bulk Murchison$^c$ | | 580 | 1700 | 10100 | 0.49 | 0.54 ± 0.08 | -0.05 ± 0.03 | 3.06 ± 0.09 | 0.34 ± 0.38 | |

$^a$ $\varepsilon^iTi = [(^iTi/^{47}Ti)_{sample}/(^iTi/^{47}Ti)_{OL-Ti} - 1] \times 10^4$ after mass-bias correction by internal normalization to $^{49}Ti/^{47}Ti = 0.749766$ using the exponential law. Uncertainties represent Student-t 95% confidence intervals $\sigma t_{0.95,n-1}/\sqrt{n}$.

$^b$ $\varepsilon^iSr = [(^iSr/^{86}Sr)_{sample}/(^iSr/^{86}Sr)_{NBS\ 987} - 1] \times 10^4$ after mass-bias correction by internal normalization to $^{86}Sr/^{88}Sr = 0.1194$ using the exponential law. Uncertainties represent $2\sigma_{SE,mean}$.

$^c$ Ti, Rb, and Sr concentration from Wasson and Kallemeyn (1988); isotope data from Trinquier et al. (2009) and Moynier et al. (2012).

Table 4 Titanium and Sr isotope data of Allende CAIs.

| CAI | Petrographic/chemical group | Rb [µg/g] ± 2σ | Sr [µg/g] ± 2σ | $^{87}$Rb/$^{86}$Sr ± 2σ | $\varepsilon^{46}$Ti$^a$ ± 2σ | $\varepsilon^{48}$Ti$^a$ ± 2σ | $\varepsilon^{50}$Ti$^a$ ± 2σ | $\varepsilon^{84}$Sr$^b$ ± 2σ | $^{87}$Sr/$^{86}$Sr ± 2σ |
|---|---|---|---|---|---|---|---|---|---|
| A-ZH-1 | Coarse-grained Type B/group I | | | | 1.55 ± 0.11 | 0.49 ± 0.06 | 9.12 ± 0.13 | 1.44 ± 0.11 | 0.700674 ± 2E-06 |
| A-ZH-2 | Coarse-grained Type B/group I | *0.490 ±* | *186.25 ±* | *0.00760 ± 0.00005* | 1.60 ± 0.07 | 0.53 ± 0.04 | 9.38 ± 0.11 | *1.18 ± 0.16* | *0.699478 ± 3E-06* |
| A-ZH-3 | Coarse-grained Type B/group I | 1.95 ± 0.18 | 195 ± 29 | 0.0290 ± 0.0051 | 1.73 ± 0.50 | 0.39 ± 0.14 | 9.38 ± 0.47 | | |
| A-ZH-4 | Coarse-grained Type B/group I | *0.70 ±* | *137.10 ±* | *0.01484 ± 9E-05* | 1.48 ± 0.06 | 0.50 ± 0.05 | 9.31 ± 0.09 | *1.39 ± 0.19* | *0.699821 ± 3E-06* |
| A-ZH-5 | Fine-gr.-Fluffy type A compound/gr. III | 3.96 ± 0.36 | 35.0 ± 5.3 | 0.327 ± 0.057 | 1.57 ± 0.10 | 0.51 ± 0.04 | 9.51 ± 0.07 | 1.07 ± 0.74 | 0.718868 ± 6E-06 |

$^a$ $\varepsilon^i$Ti = [($^i$Ti/$^{47}$Ti)$_{sample}$/($^i$Ti/$^{47}$Ti)$_{OL-Ti}$ − 1] × 10$^4$ after mass-bias correction by internal normalization to $^{49}$Ti/$^{47}$Ti = 0.749766 using the exponential law. Uncertainties represent Student-t 95% confidence intervals $\sigma t_{0.95,n-1}/\sqrt{n}$.

$^b$ $\varepsilon^i$Sr = [($^i$Sr/$^{86}$Sr)$_{sample}$/($^i$Sr/$^{86}$Sr)$_{NBS\ 987}$ − 1] × 10$^4$ after mass-bias correction by internal normalization to $^{86}$Sr/$^{88}$Sr = 0.1194 using the exponential law. Uncertainties represent 2$\sigma_{SE,mean}$.

Values in italic are from Hans et al. (2013) and have been obtained after mild leaching of the CAIs in HCl to remove secondary Rb.

Table 5 Concentration data used for producing Figs. 7, 9, 10

| | Reservoir | Si | Ca | Ti | Cr | Ni | Sr | Zr | Mo | Ru | Ba | Nd | Sm | Hf | W | Os |
|---|---|---|---|---|---|---|---|---|---|---|---|---|---|---|---|---|
| CI | CC | 107000 | 9220 | 458 | 2650 | 10800 | 7.3 | 3.6 | 1.0 | 0.69 | 2.4 | 0.47 | 0.15 | 0.12 | 0.10 | 0.49 |
| CM | CC | 129000 | 12700 | 634 | 3050 | 12000 | 10.1 | 8.0 | 1.3 | 0.88 | 3.3 | 0.63 | 0.20 | 0.19 | 0.14 | 0.64 |
| CO | CC | 159000 | 15800 | 780 | 3550 | 14000 | 12.7 | 7.8 | 1.5 | 1.09 | 4.3 | 0.77 | 0.24 | 0.18 | 0.16 | 0.79 |
| CV | CC | 159600 | 19000 | 897 | 3600 | 13400 | 16.0 | 8.3 | 1.5 | 1.13 | 4.9 | 0.97 | 0.31 | 0.19 | 0.19 | 0.83 |
| OC mean | NC | 175000 | 13000 | 620 | 3700 | 14000 | 10.2 | 6.0 | 1.1 | 0.90 | 4.2 | 0.65 | 0.19 | 0.17 | 0.14 | 0.60 |
| EC mean | NC | 175000 | 9200 | 510 | 3060 | 15250 | 7.7 | 5.0 | 1.0 | 0.88 | 2.6 | 0.49 | 0.14 | 0.15 | 0.13 | 0.63 |
| Av. CV3 CAIs | CAI | 125600 | 164300 | 6042 | 200 | 900 | 100 | 40 | 3.5 | 5 | 30 | 14 | 4.54 | 1.2 | 1 | 9 |

Data sources are Wasson and Kallemeyn (1988), Stracke et al. (2012), Kleine et al. (2004), Burkhardt et al. (2008;2011;2016), Trinquier et al. (2009), Archer et al. (2014), and therein. Note that the elemental composition of different types of CAIs is variable (Stracke et al. 2012) and that the best estimates of the average CV3 CAI composition given here depend on the mix of inclusions.

Table 6 Compilation of isotope anomalies in bulk planetary materials, average CV3 CAIs, and the weighted average of Murchison leachates.

| | Reservoir | $\varepsilon^{48}$Ca | 2σ | $\varepsilon^{50}$Ti | 2σ | $\varepsilon^{54}$Cr | 2σ | $\varepsilon^{62}$Ni | 2σ | $\varepsilon^{84}$Sr | 2σ | $\varepsilon^{96}$Zr | 2σ | $\varepsilon^{94}$Mo | 2σ | $\varepsilon^{100}$Ru | 2σ | $\varepsilon^{135}$Ba | 2σ | $\varepsilon^{145}$Nd | 2σ | $\varepsilon^{144}$Sm | 2σ | $\varepsilon^{180}$Hf | 2σ | $\varepsilon^{183}$W | 2σ | $\varepsilon^{186}$Os | 2σ |
|---|---|---|---|---|---|---|---|---|---|---|---|---|---|---|---|---|---|---|---|---|---|---|---|---|---|---|---|---|---|
| **Carbonaceous chond.** | | | | | | | | | | | | | | | | | | | | | | | | | | | | | |
| CI | CC | 2.05 | 0.20 | 1.85 | 0.12 | 1.56 | 0.06 | 0.20 | 0.14 | 0.40 | 0.10 | 0.34 | 0.24 | 0.79 | 0.41 | −0.24 | 0.13 | 0.27 | 0.14 | −0.21 | 0.13 | −1.02 | 0.46 | | | | | 0.13 | 0.17 |
| CM | CC | 3.14 | 0.14 | 3.01 | 0.10 | 1.02 | 0.08 | 0.10 | 0.03 | 0.40 | 0.09 | 0.76 | 0.37 | 4.03 | 1.43 | −1.23 | 1.51 | 0.34 | 0.31 | 0.07 | 0.06 | −0.65 | 0.66 | −0.11 | 0.46 | 0.09 | 0.24 | 0.46 | 0.53 |
| CO | CC | 3.87 | 0.56 | 3.77 | 0.50 | 0.77 | 0.33 | 0.11 | 0.03 | 0.41 | 0.21 | 0.69 | 0.25 | 1.29 | 0.40 | −0.92 | 1.22 | | | | | | | 0.01 | 0.40 | 0.02 | 0.15 | | |
| CV | CC | 3.92 | 0.50 | 3.49 | 0.20 | 0.87 | 0.07 | 0.11 | 0.03 | 0.63 | 0.10 | 1.10 | 0.31 | 1.47 | 0.45 | −1.17 | 0.22 | 0.26 | 0.41 | 0.03 | 0.05 | −0.87 | 0.17 | 0.05 | 0.30 | 0.04 | 0.04 | −0.24 | 0.14 |
| CK | CC | | | 3.63 | 0.40 | 0.48 | 0.30 | | | | | 0.45 | 0.25 | 1.55 | 0.36 | −1.10 | 0.23 | | | | | | | 0.21 | 0.37 | −0.07 | 0.17 | −0.80 | 0.29 |
| CR | CC | 2.15 | 0.11 | 2.74 | 1.17 | 1.31 | 0.10 | 0.09 | 0.05 | | | 1.03 | 0.40 | 2.82 | 0.38 | −0.76 | 0.45 | | | | | | | 0.38 | 0.19 | | | | |
| CH | CC | | | | | 1.37 | 0.29 | | | | | | | | | −0.91 | 0.13 | | | | | | | 0.12 | 0.17 | | | | |
| CB | CC | | | 2.04 | 0.07 | 1.21 | 0.09 | | | | | 0.96 | 0.25 | 1.30 | 0.26 | −1.04 | 0.13 | | | | | | | 0.10 | 0.15 | | | | |
| **Ordinary chondrites** | | | | | | | | | | | | | | | | | | | | | | | | | | | | | |
| H | NC | −0.24 | 0.30 | −0.64 | 0.17 | −0.36 | 0.08 | −0.06 | 0.03 | −0.13 | 0.10 | 0.67 | 0.44 | 0.72 | 0.24 | −0.27 | 0.04 | 0.08 | 0.13 | 0.05 | 0.03 | 0.06 | 0.06 | 0.11 | 0.40 | | | 0.18 | 0.12 |
| L | NC | −0.32 | 0.03 | −0.63 | 0.08 | −0.40 | 0.10 | −0.04 | 0.03 | −0.11 | 0.21 | | | 0.66 | 0.39 | −0.28 | 0.13 | 0.14 | 0.04 | 0.09 | 0.13 | 0.02 | 0.17 | 0.02 | 0.30 | | | | |
| LL | NC | −0.15 | 0.39 | −0.67 | 0.08 | −0.42 | 0.07 | −0.07 | 0.03 | −0.43 | 0.10 | 0.34 | 0.25 | 0.74 | 0.72 | 0.12 | 0.70 | 0.08 | 0.05 | 0.06 | 0.18 | 0.05 | 0.23 | −0.1 | 0.27 | | | | |
| OC Mean | NC | −0.28 | 0.20 | −0.65 | 0.07 | −0.39 | 0.04 | −0.06 | 0.02 | −0.17 | 0.15 | 0.60 | 0.35 | 0.70 | 0.14 | −0.24 | 0.06 | 0.10 | 0.05 | 0.06 | 0.03 | 0.05 | 0.05 | 0.07 | 0.20 | 0.09 | 0.17 | 0.18 | 0.12 |
| **Rumuruti chondrites** | | | | | | | | | | | | | | | | | | | | | | | | | | | | | |
| R | NC | | | | | 0.16 | 0.50 | | | | | | | | | −0.39 | 0.13 | | | | | | | −0.35 | 0.50 | −0.01 | 0.18 | | |
| **Enstatite chondrites** | | | | | | | | | | | | | | | | | | | | | | | | | | | | | |
| EH | NC | −0.32 | 0.56 | −0.14 | 0.07 | 0.04 | 0.08 | 0.03 | 0.03 | −0.21 | 0.12 | −0.02 | 0.30 | 0.59 | 0.36 | −0.16 | 0.25 | 0.17 | 0.07 | 0.05 | 0.03 | −0.02 | 0.13 | | | −0.24 | 0.48 | | |
| EL | NC | −0.38 | 0.26 | −0.28 | 0.17 | 0.04 | 0.07 | −0.03 | 0.07 | | | | | 0.36 | 0.09 | −0.08 | 0.05 | | | 0.03 | 0.04 | 0.27 | 0.24 | | | −0.38 | 0.41 | | |
| EC Mean | NC | −0.37 | 0.46 | −0.20 | 0.08 | 0.04 | 0.05 | 0.00 | 0.07 | −0.21 | 0.12 | −0.02 | 0.30 | 0.41 | 0.08 | −0.08 | 0.03 | 0.17 | 0.07 | 0.04 | 0.02 | 0.00 | 0.08 | −0.05 | 0.15 | −0.31 | 0.44 | | |
| Av. CV3 CAIs | CAI | 3.80 | 1.33 | 8.77 | 0.53 | 7.00 | 1.00 | 1.00 | 0.30 | 1.26 | 0.11 | 1.61 | 0.31 | 1.23 | 0.19 | −1.60 | 0.07 | 0.54 | 0.06 | −0.23 | 0.03 | −2.34 | 0.10 | 0.30 | 0.02 | 0.08 | 0.14 | 0.50 | 0.70 |
| Wt. av. Murch. leach. | CC | 3.25 | 0.23 | 3.10 | 0.17 | 1.63 | 0.10 | | | 0.35 | 0.18 | 1.03 | 0.68 | 3.37 | 0.30 | | | 0.61 | 0.29 | 0.05 | 0.06 | −0.7 | 0.30 | −0.57 | 0.22 | 0.12 | 0.42 | 1.65 | 1.44 |
| **Achondrites** | | | | | | | | | | | | | | | | | | | | | | | | | | | | | |
| Acapulcoites | NC | | | −1.30 | 0.05 | −0.75 | 0.25 | | | | | | | | | | | | | | | | | | | −0.06 | 0.17 | | |
| Lodranites | NC | | | | | −0.34 | 0.25 | | | | | | | 1.10 | 0.30 | | | | | | | | | | | −0.09 | 0.13 | | |
| Brachinites | NC | | | | | | | | | | | | | | | | | | | | | | | | | | | | |
| Winonaites | NC | −0.21 | 0.09 | | | | | | | | | | | 0.21 | 0.30 | | | | | | | | | | | | | | |
| NWA5363/5400 | NC | −0.53 | 0.20 | −1.01 | 0.10 | −0.37 | 0.13 | 0.01 | 0.03 | | | | | 0.69 | 0.19 | −0.34 | 0.13 | | | 0.11 | 0.06 | 0.27 | 0.43 | | | | | | |
| Angrites | NC | −1.06 | 0.33 | −1.18 | 0.08 | −0.43 | 0.15 | 0.01 | 0.03 | 0.00 | 0.10 | | | 0.70 | 0.50 | | | | | −0.11 | 0.26 | | | | | −0.06 | 0.09 | | |
| Aubrites | NC | −0.44 | 0.59 | −0.06 | 0.11 | −0.16 | 0.19 | 0.05 | 0.19 | | | | | | | | | | | | | | | | | | | | |
| HED | NC | −1.24 | 0.29 | −1.23 | 0.05 | −0.67 | 0.09 | 0.03 | 0.12 | 0.01 | 0.10 | 0.47 | 0.21 | | | | | | | 0.05 | 0.04 | 0.10 | 0.04 | | | −0.03 | 0.13 | | |
| Urelites | NC | −1.82 | 0.44 | −1.85 | 0.26 | −0.92 | 0.05 | −0.04 | 0.10 | | | | | | | | | | | | | | | | | 0.00 | 0.04 | | |
| Mars' mantle | NC | −0.20 | 0.03 | −0.43 | 0.06 | −0.18 | 0.05 | 0.04 | 0.03 | −0.25 | 0.15 | | | 0.37 | 0.10 | | | | | −0.02 | 0.01 | | | | | 0.02 | 0.13 | | |
| Moon | NC | 0.04 | 0.03 | −0.03 | 0.04 | 0.15 | 0.10 | | | −0.22 | 0.13 | | | | | | | | | | | | | | | −0.01 | 0.15 | | |
| Earth mantle | | 0 | 0 | 0 | 0 | 0 | 0 | 0 | 0 | −0.19 | 0.16 | 0.06 | 0.04 | 0 | 0 | 0 | 0 | 0 | 0 | 0 | 0 | 0 | 0 | 0 | 0 | 0 | 0 | 0 | 0 |
| Mesosiderites | NC | | | −1.27 | 0.16 | −0.69 | 0.11 | | | | | | | | | | | | | | | | | | | | | | |
| **Pallasites** | | | | | | | | | | | | | | | | | | | | | | | | | | | | | |
| MG pal | NC | | | −1.37 | 0.08 | −0.39 | 0.4 | | | | | | | 0.85 | 0.22 | −0.45 | 0.29 | | | | | | | | | | | ~0 | 0.2 |
| Eagle station Pal | CC | | | | | 0.7 | 0.1 | | | | | | | 0.85 | 0.32 | | | | | | | | | | | | | ~0 | 0.2 |
| **Iron meteorites** | | | | | | | | | | | | | | | | | | | | | | | | | | | | | |
| IAB | NC | | | | | | | | | | | | | 0.00 | 0.07 | −0.06 | 0.07 | | | | | | | | | −0.06 | 0.04 | ~0 | 0.2 |
| IC | NC | | | | | | | | | | | | | 0.97 | 0.33 | | | | | | | | | | | −0.05 | 0.06 | | |
| IIAB | NC | | | | | −0.10 | 0.12 | | | | | | | 1.08 | 0.18 | −0.46 | 0.05 | | | | | | | | | −0.02 | 0.02 | ~0 | 0.2 |
| IIC | CC | | | | | −0.10 | 0.09 | | | | | | | 2.27 | 0.14 | | | | | | | | | | | 0.28 | 0.07 | | |
| IID | CC | | | | | | | | | | | | | 1.07 | 0.26 | −0.74 | 0.48 | | | | | | | | | 0.11 | 0.07 | | |
| IIE | NC | | | | | −0.59 | 0.13 | | | | | | | 0.64 | 0.20 | | | | | | | | | | | | | | |
| IIF | CC | | | | | | | | | | | | | 1.11 | 0.13 | | | | | | | | | | | 0.09 | 0.02 | | |
| IIIAB | NC | | | | | −0.85 | 0.06 | −0.12 | 0.02 | | | | | 0.98 | 0.20 | −0.63 | 0.06 | | | | | | | | | −0.03 | 0.02 | ~0 | 0.2 |
| IIICD | NC | | | | | | | | | | | | | −0.06 | 0.32 | | | | | | | | | | | | | | |
| IIIE | NC | | | | | | | | | | | | | 0.91 | 0.08 | | | | | | | | | | | −0.05 | 0.04 | | |
| IIIF | CC | | | | | | | | | | | | | 1.13 | 0.31 | | | | | | | | | | | 0.08 | 0.07 | | |
| IVA | NC | | | | | −0.07 | 0.05 | | | | | | | 0.68 | 0.17 | −0.29 | 0.09 | | | | | | | | | −0.03 | 0.07 | ~0 | 0.2 |
| IVB | CC | | | | | 0.06 | 0.04 | | | | | | | 1.38 | 0.14 | −0.90 | 0.05 | | | | | | | | | 0.13 | 0.02 | ~0 | 0.2 |

Classification of meteorites into NC and CC is based on isotopic dichotomy in Cr, Ti, and Mo isotopes (Warren 2011; Trinquier et al., 2007; Leya et al., 2008; Budde et al. 2016). Uncertainties represent Student-t 95% confidence intervals $\sigma t_{0.95,n-1}/\sqrt{n}$. Note that for some elements (*e.g.*, Ti; *cf.* Fig. 6) the isotopic composition measured for individual CAIs spans a significantly larger range than what is indicated by the 95%ci. This suggests admixing of NC material to CAIs at the level of individual inclusions, or spacial/temporal heterogeneities in the CAI forming reservoir. Data sources as in compilation by Burkhardt et al. (2017), with addition of Brennecka et al. (2013), Akram et al. (2013), Huang et al. (2012), Schiller et al. (2018), Davis et al. (2018), Birck (2004), Kruijer et al. (2017), Worsham et al. (2017), Budde et al. (2018), Walker (2012), Sprung et al. (2010), Yokoyama et al. (2007;2010), and therein.

**Figure captions**

**Figure 1.:** Nonexclusive scenarios proposed to account for the isotopic heterogeneity at the bulk meteorite scale. Planetary-scale isotope anomalies are either (a) a primordial feature of the solar nebula inherited from heterogeneities in the infalling presolar molecular cloud materials, or (b) they are caused by fractionation processes within the solar nebula, such as separation of dust grains of different types and sizes, or the selective destruction of thermally labile presolar components.

**Figure 2.:** Release patterns of selected elements obtained by a six-step sequential digestion of a powdered sample of the Murchison (CM2) meteorite.

**Figure 3.:** Chondrite-normalized REE abundance patterns. (a) Murchison leachates exhibit a large range in concentrations and subtle differences in the shape of the patterns, however, the cumulative total of all leachates and the residue has a flat REE pattern and an enrichment (~1.25×CI) consistent with bulk CM literature data. (b) Investigated Allende CAIs exhibit about flat REE patterns with ~15×CI enrichment. CAI A-ZH-5 has negative volatility controlled excursions in Eu and Yb (group III pattern).

**Figure 4.:** Titanium and Sr isotopic anomaly data of Murchison leachates and Allende CAIs. (a,b) The leachate samples show variable nucleosynthetic Ti and Sr isotopic anomalies, the weighted average of which is consistent with bulk Murchison literature data. (c,d) Anomalies in the CAIs are homogeneous.

**Figure 5.:** $^{87}$Rb-$^{87}$Sr isochron diagrams for Murchison leachates and Allende CAIs. a) Individual leach steps show no isochronous relationship, while the weighted average of the leachates plots on a 4.567 Ga chondrite isochron. This indicates closed system evolution on the the Murchison parent body, but incongruent dissolution of Rb and Sr during the leaching procedure. b.) Allende CAIs for which Rb/Sr and Sr isotopic composition data is available fall within uncertainty on a 4.567 Ga chondrite isochron.

**Figure 6.:** Titanium isotope composition of Murchison leachates and various planetary materials in $\varepsilon^{46}$Ti, $\varepsilon^{48}$Ti, and $\varepsilon^{50}$Ti space. Anomalies vary by three orders of magnitude from presolar grains (panels g, h, i) via hibonites (panels d, e, f) to CAIs, individual chondrules, and bulk planetary bodies (panels a, b, c). For clarity, error bars of SiC grains were omitted in panels d, e, f.

**Figure 7.:** Mixing diagrams of CAIs and average ordinary (OC) and enstatite chondrites (EC) for $\varepsilon^{50}$Ti vs. Si/Ti and $\varepsilon^{84}$Sr vs. Si/Sr. Carbonaceous chondrites (CC) fall along or between mixing trajectories. This indicates that the presence of several percent (by mass) of CAI-like refractory materials in CC bodies can satisfy both, the refractory element enrichment of CC relative to NC bodies, as well as the offset and variation in nucleosynthetic anomalies. Data used to construct diagrams are given in Tables 5 and 6.

**Figure 8.:** $\varepsilon^{84}$Sr vs. $\varepsilon^{94}$Mo of leachates is positively correlated, and roughly follow the expected co-variation resulting from a variable distribution of *s*-process nuclides (dashed line, calculated using the formalism of Dauphas et al. (2004) and the *s*-process abundances of Bisterzo et al. (2014) and Arlanini et al. (1999). The acetic acid leachate (L1) plots significantly off the calculated $\varepsilon^{84}$Sr–$\varepsilon^{92}$Mo correlation line, indicating fluid-assisted Sr mobilization and redistribution on the parent body.

**Figure 9.:** $\varepsilon^{50}$Ti vs. $\varepsilon^{48}$Ca, $\varepsilon^{54}$Cr, $\varepsilon^{84}$Sr, $\varepsilon^{96}$Zr, $\varepsilon^{94}$Mo, $\varepsilon^{100}$Ru, $\varepsilon^{135}$Ba, $\varepsilon^{145}$Nd, $\varepsilon^{144}$Sm, $\varepsilon^{180}$Hf, $\varepsilon^{183}$W, and $\varepsilon^{186}$Os isotope anomalies in leachates and acid residues of the Murchison meteorite along with bulk NC, CC and CAI data, mixing lines between OC (purple crosses), EC (orange crosses) and CAIs, and approximate trajectories of s-process mixing lines. The Ca, Ti, Cr, Sr, Mo, Ru, W, and Os Murchison leachate data were acquired from the same 6-step sequential digestion, Zr data are from a 5-step digestion (Schönbächler et al. 2005), and Ba, Nd, Sm, and Hf data are from a 4-step digestion (Qin et al. 2011). Titanium isotope anomalies of the leachates are well correlated with geochemically similar elements like Zr, Hf, Nd, as well as with Mo and W. The general trend is from a *s*-deficit in L1 to a *s*-excess in L6. Very poor, if any, correlation exists in the leachates for Ti and Ca, Cr, Ru, and Os although some of these elements like Ca and Ti show tight correlation at a bulk meteorite scale. The reason for the decoupling reflects that fact that these elements have different geochemical behaviors so that mixing between different mineral fractions tapped by the various leaching steps follow complex topologies in mixing spaces. The weighted average anomaly of the leachates is well within the range of bulk CC bodies for all elements analyzed. For no element pair the offset between NC and CC meteorites is clearly attributable to the correlations seen in the leachates, highlighting that the offset is decoupled from the dominant intrinsic anomalies as recorded in the leachates. For data sources see main text and Tables 5 and 6. Equivalent diagrams with $\varepsilon^{84}$Sr and $\varepsilon^{94}$Mo on the abscissa are provided in supplementary Figures S1 and S2.

**Figure 10.:** Like Fig. 9 but focused on the relation of anomalies in bulk NC and CC bodies, and CAIs. For the anomaly pairs $\varepsilon^{50}$Ti vs. $\varepsilon^{84}$Sr, $\varepsilon^{96}$Zr, $\varepsilon^{135}$Ba, and $\varepsilon^{180}$Hf the offset of NC and CC and the range within each reservoir is well-explained by admixing of CAIs to EC and OC compositions. For these elements the CAI anomalies are dominant and within group *s*-process variability is minor. For Mo, Ru, Nd significant nucleosynthetic *s*-process variability within the NC and/or the CC reservoirs have been demonstrated, such that the location and range of the NC and CC meteorites in $\varepsilon^{50}$Ti vs. $\varepsilon^{94}$Mo, $\varepsilon^{100}$Ru, and $\varepsilon^{145}$Nd space is a function of both, *s*-process variability and variable addition of CAI-like material. In $\varepsilon^{50}$Ti vs. $\varepsilon^{54}$Cr, and $\varepsilon^{62}$Ni space CC bodies are offset from NC bodies towards CAIs. However, since CAIs are depleted in Cr and Ni relative to bulk chondrites, mixing lines of CAIs and EC and OC compositions do not pass through the CC bodies. This suggests that CAIs are only the refractory part of an isotopically distinct nebular reservoir and less refractory material with ~CAI-like isotopic signature also plays a role in setting planetary scale anomalies. Equivalent diagrams with $\varepsilon^{84}$Sr and $\varepsilon^{94}$Mo on the abscissa are provided in supplementary Figures S3 and S4. All panels and mixing lines can be reproduced with the data given in Tables 5 and 6.

**Figure 11.:** A qualitative model of the early Solar System that accounts for key observables in the meteoritic record. The general set-up is adopted from Yang and Ciesla (2012) and Pignatale et al. (2018), but now the collapsing cloud core contains isotopically slightly distinct parcels, one with a CAI-like isotopic composition and one with a NC-like isotopic composition. A) The initial stages of collapse are dominated by the isotopically CAI-like cloud parcel and the centrifugal radius of the infall is limited to regions close to the forming star *i.e.*, most of the material is processed at high temperatures. Through viscous spreading of the forming disk, this early infalling material is rapidly transported outwards. This process has been suggested to explain the overabundance of refractory phases in outer Solar System materials (Yang and Ciesla, 2012). B) The centrifugal radius of infalling material is expanding with time, and the primary signature of the early infall will be diluted by the addition of NC cloud material, particularly in the inner disk regions. C) Although material processing and transport in the disk will further dilute and mix the initial isotopic

signatures, the outer disk retains a higher fraction of matter from the isotopically CAI-like reservoir, and also receives a higher fraction of unprocessed primitive cloud material. D) At the end of infall the inner part of the disk contains high amounts of highly processed, volatile-poor material with an NC signature, while the outer disk has an excess in refractories from the early infall, a higher portion of unprocessed primitive dust, higher volatile contents, and an isotopic composition between NC and CAIs. This early set dichotomy was retained throughout accretion, possibly assisted by the early formation of Jupiter (Kruijer et al. 2017).

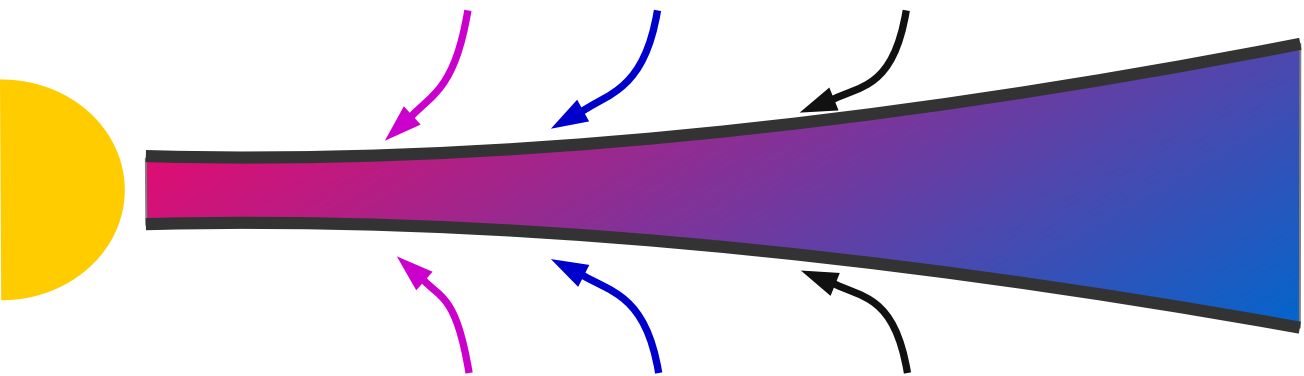

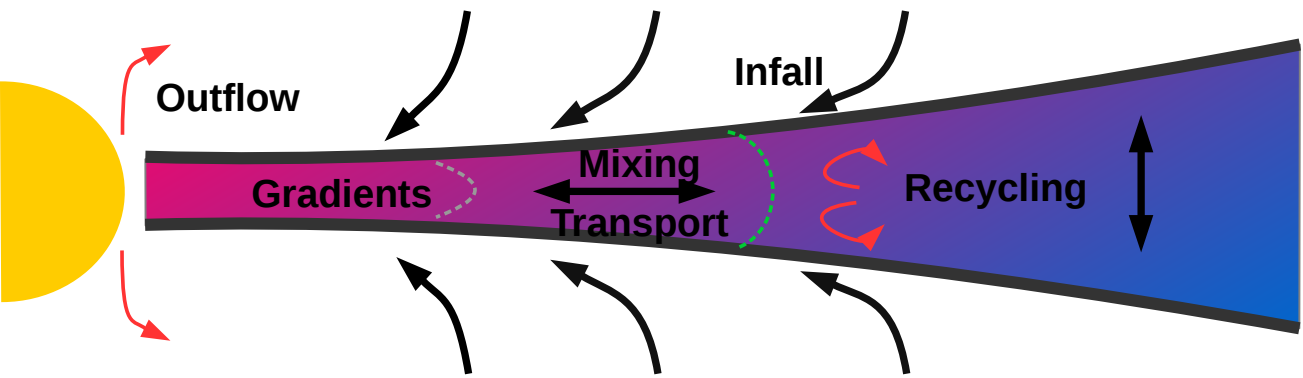

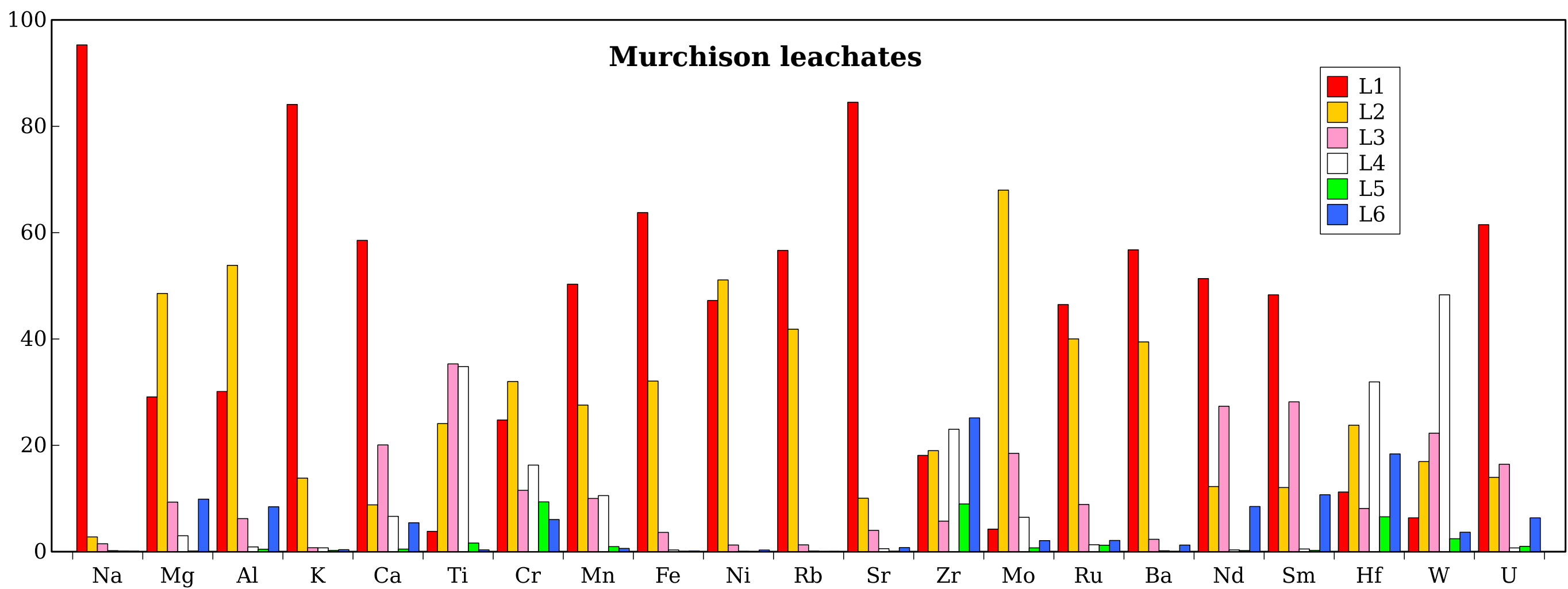

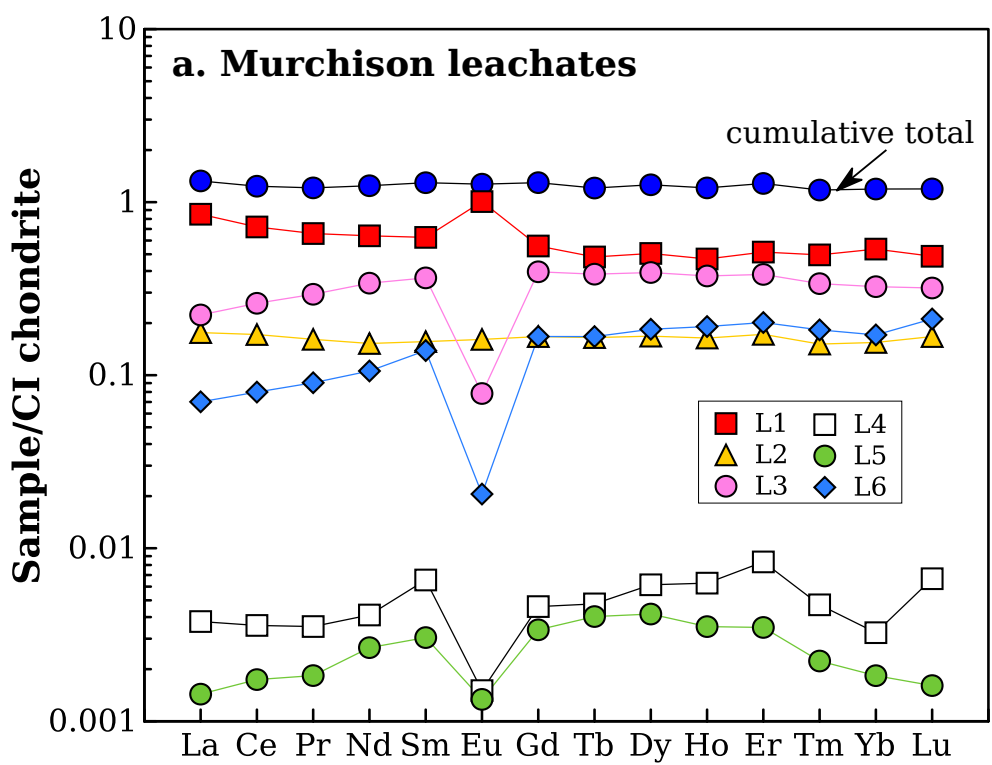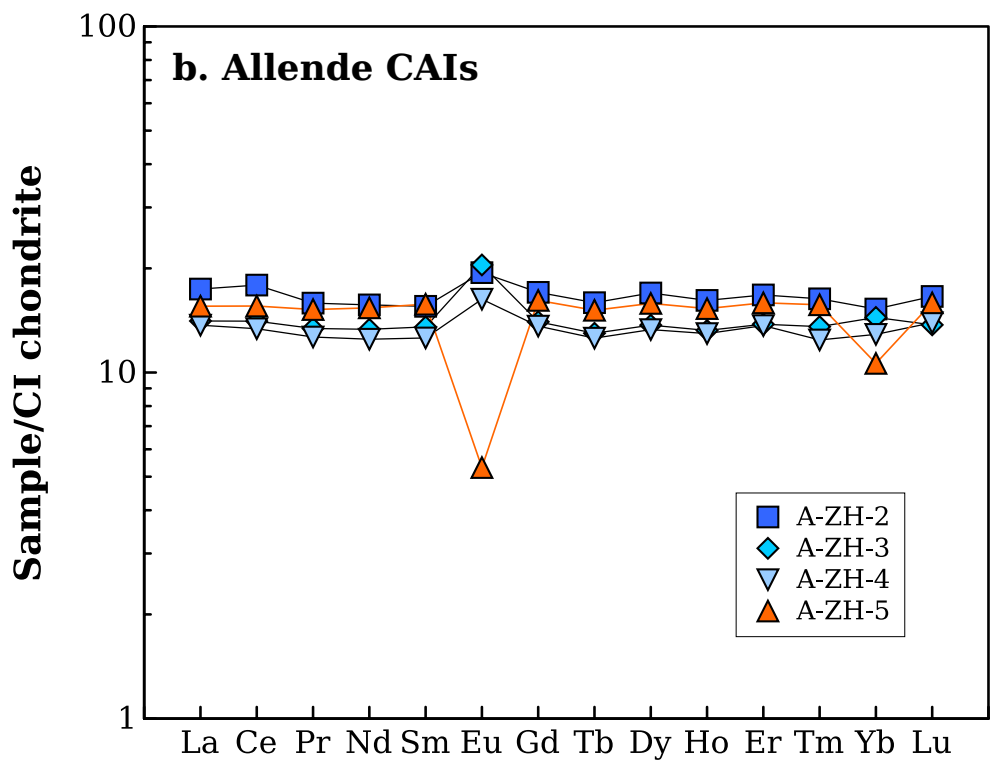

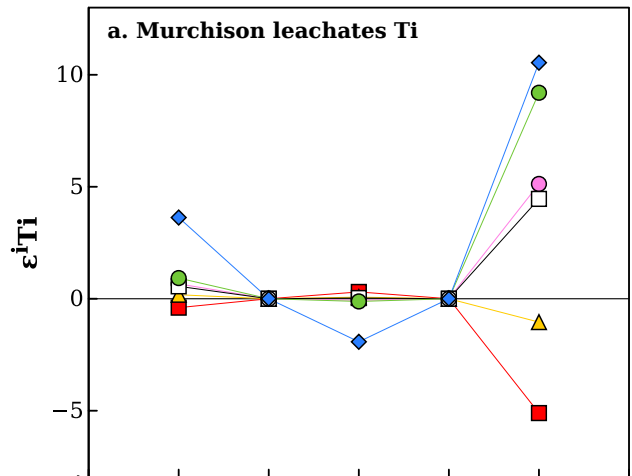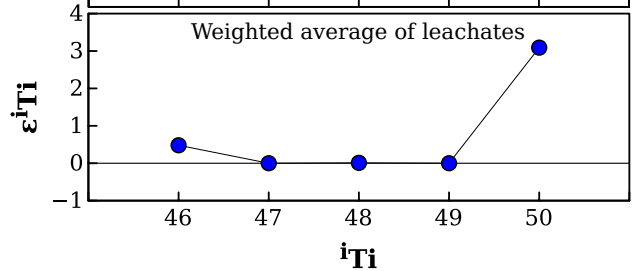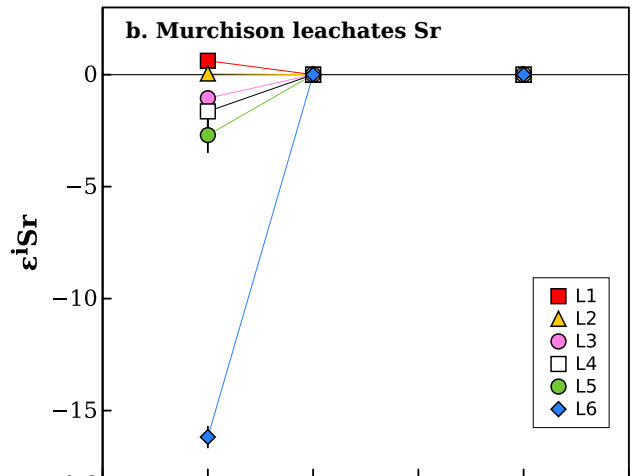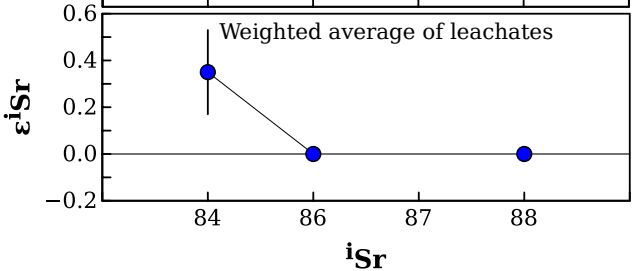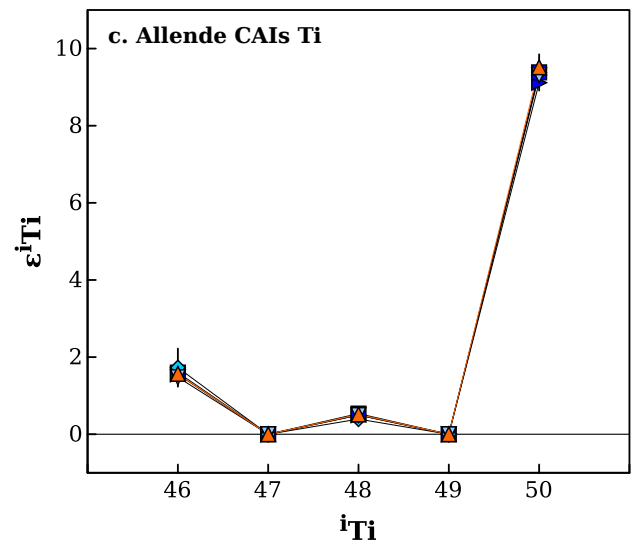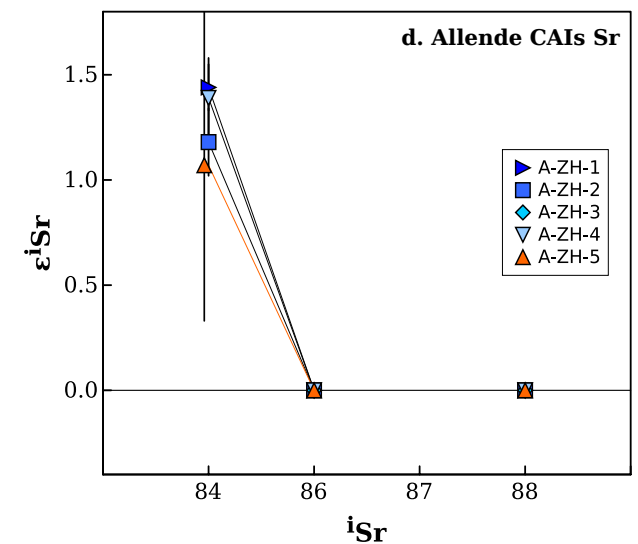

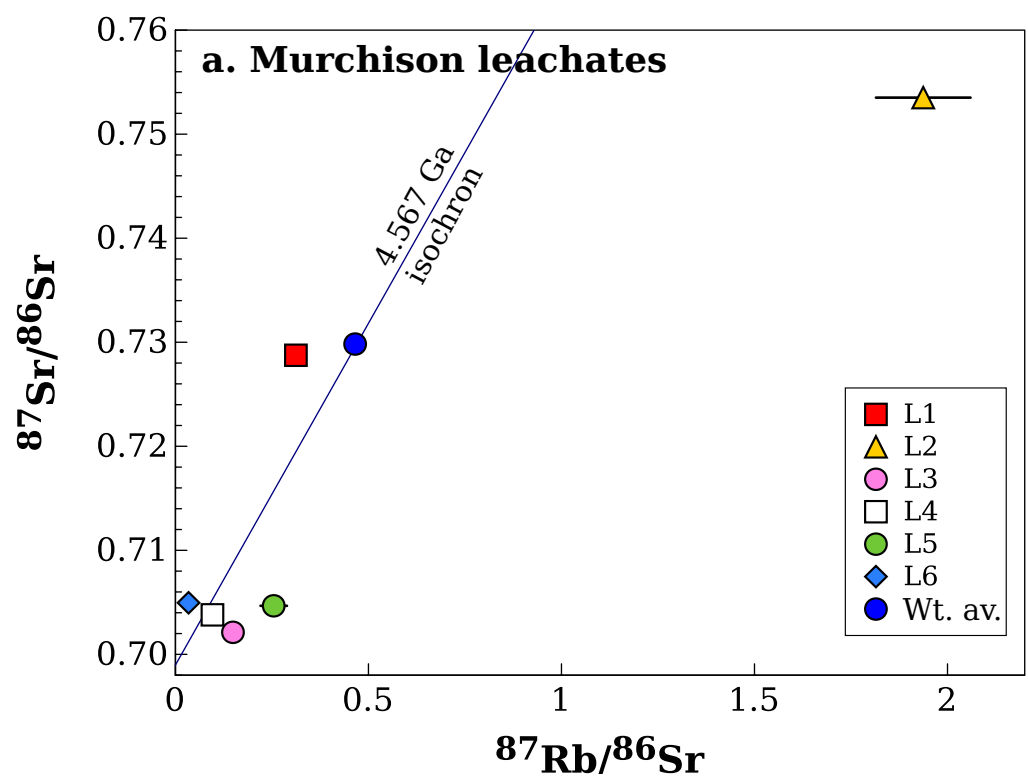
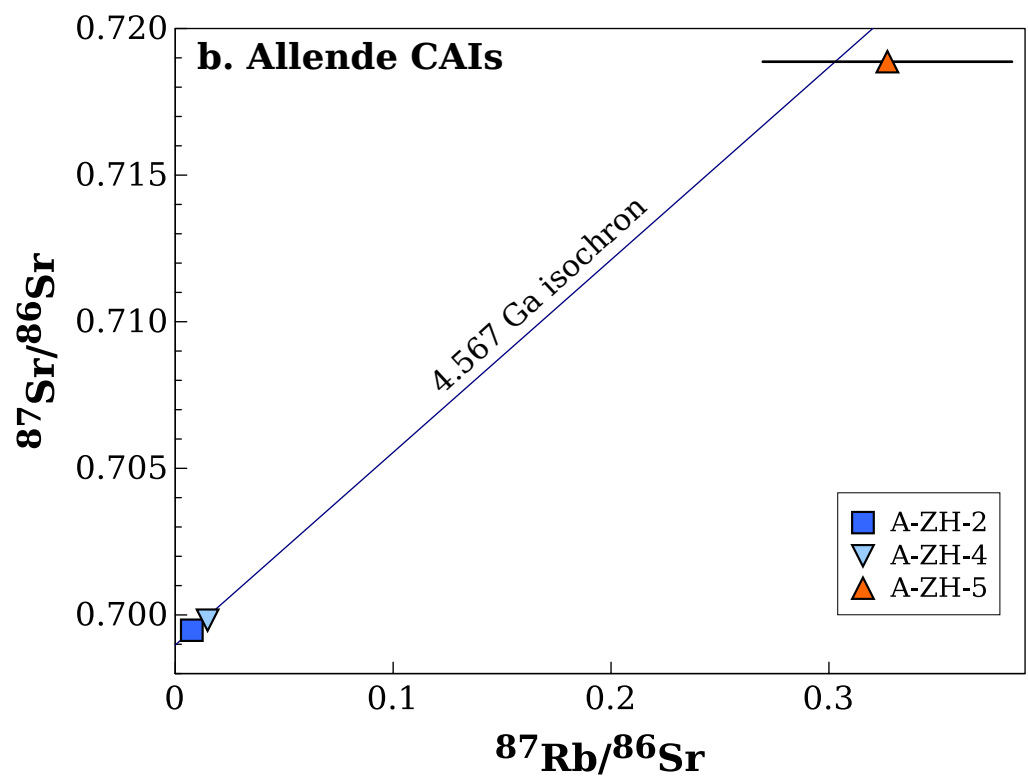

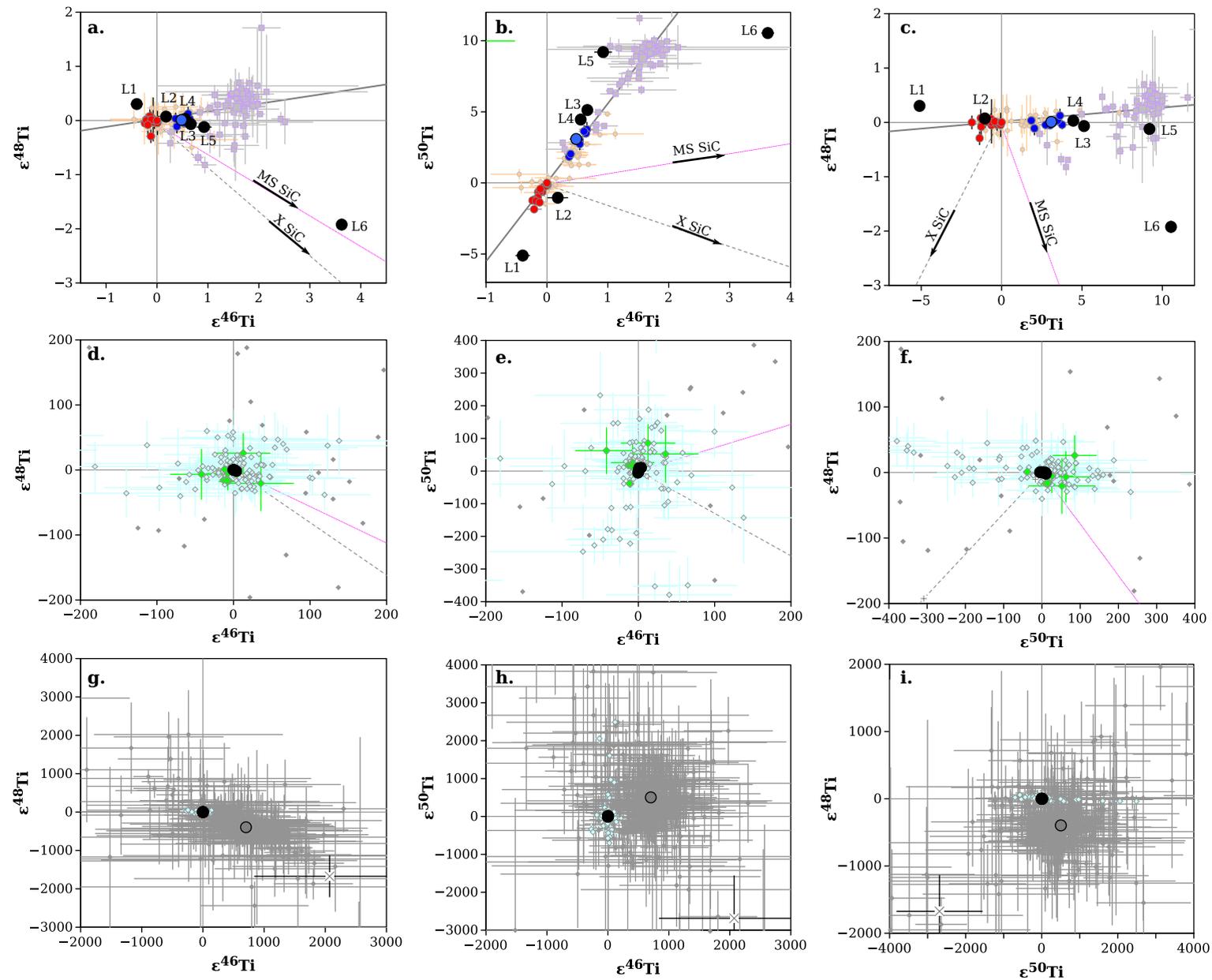

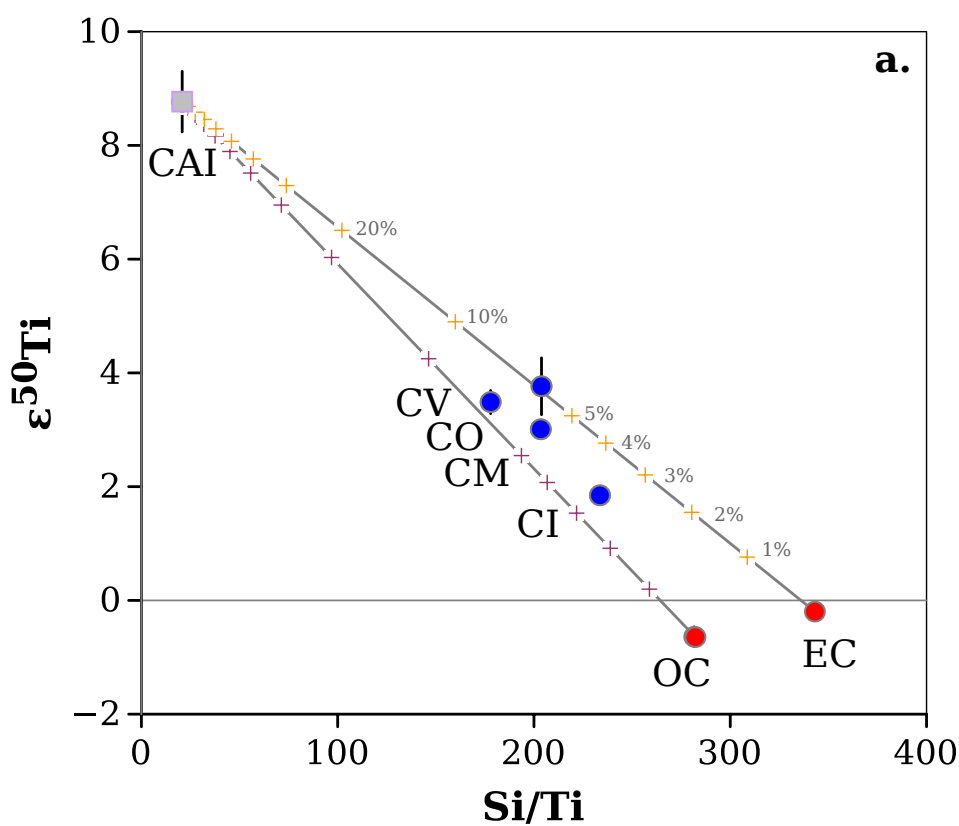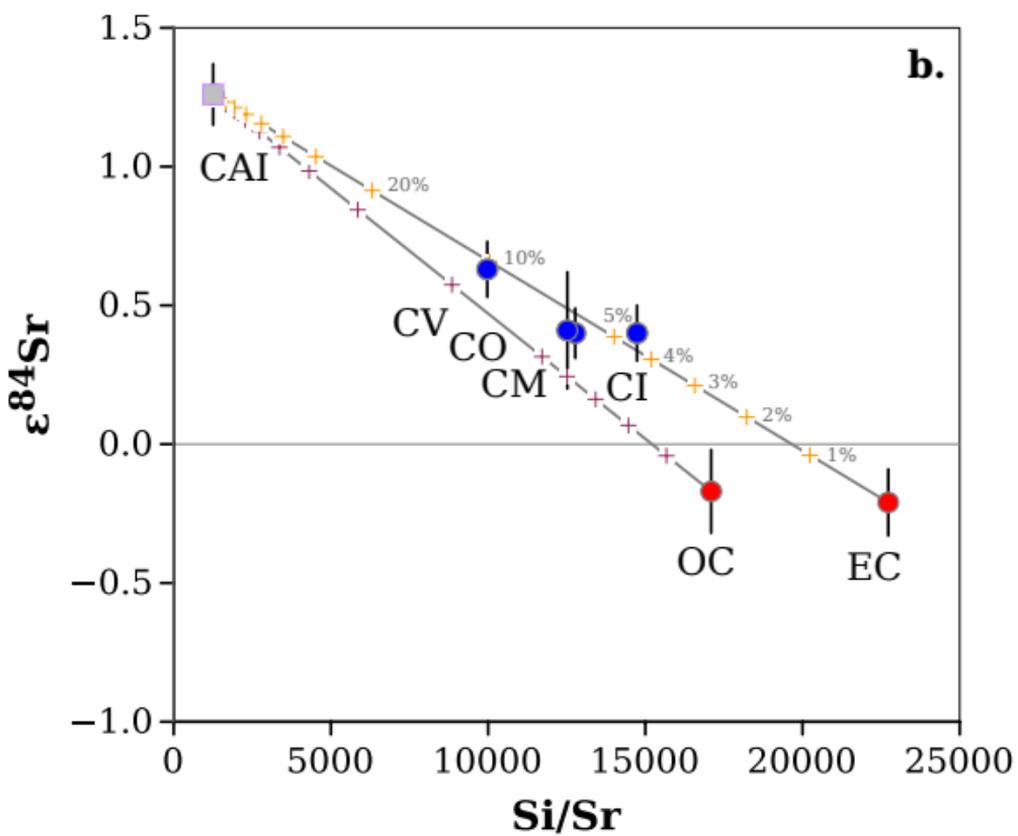

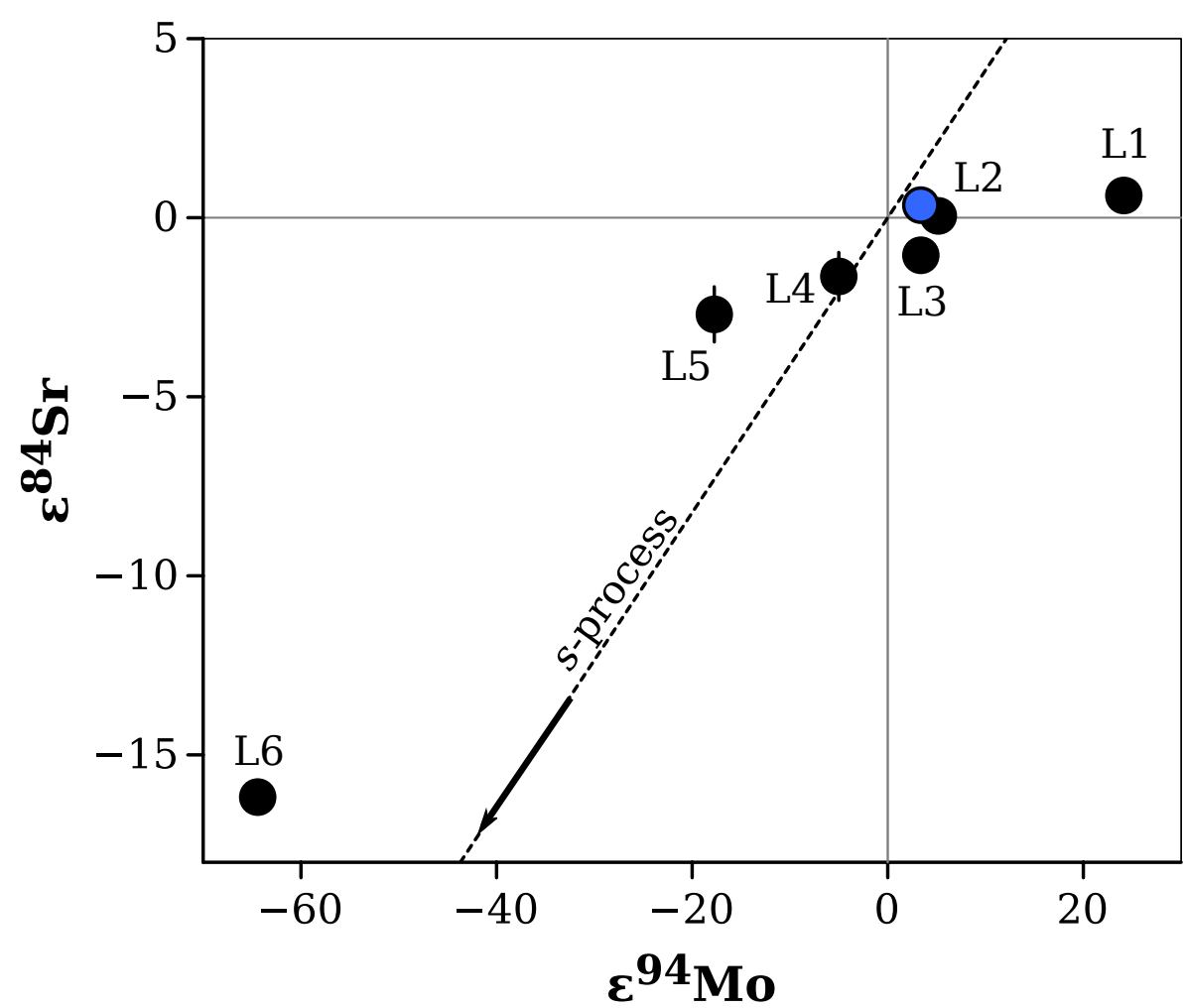

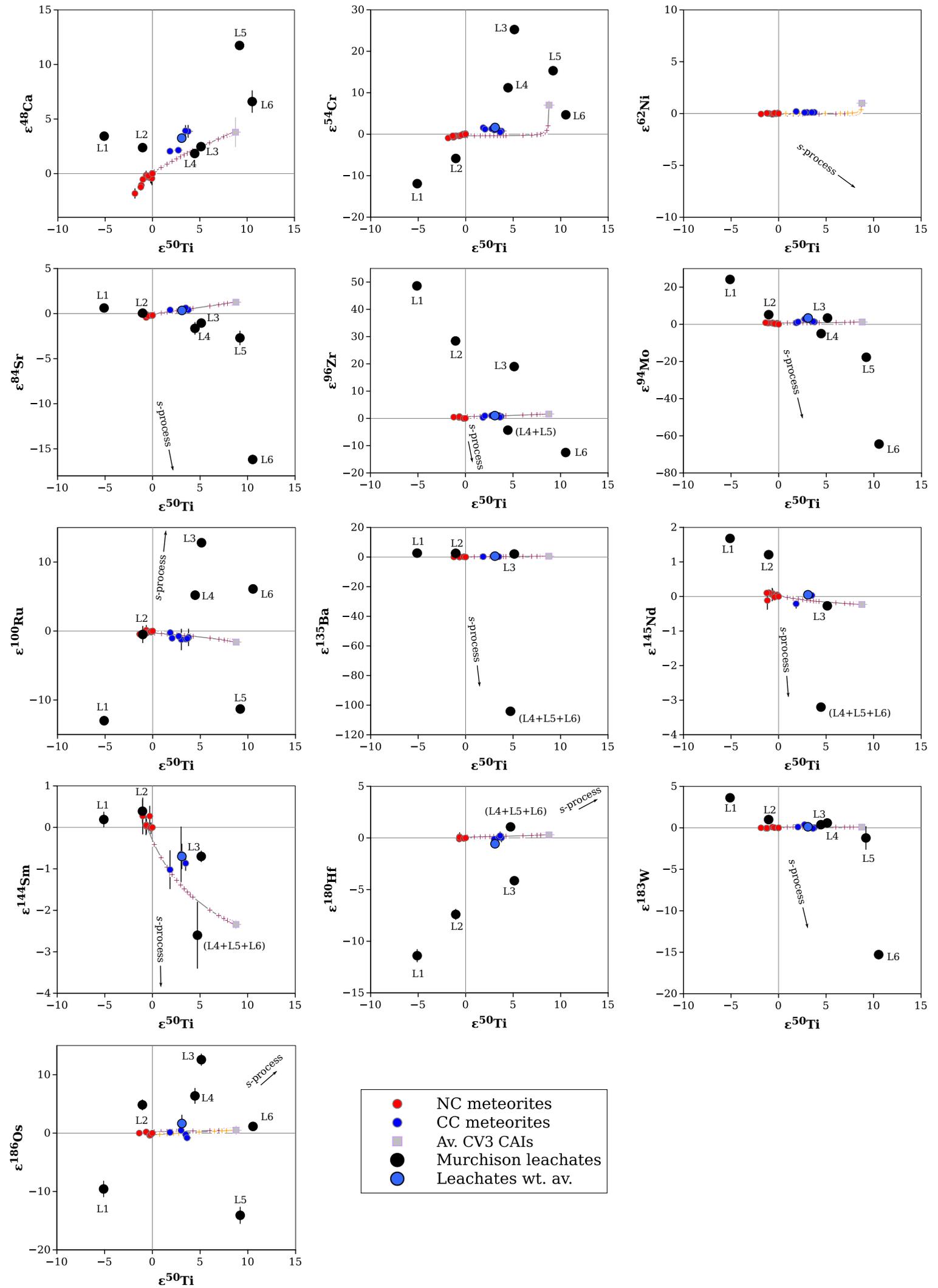

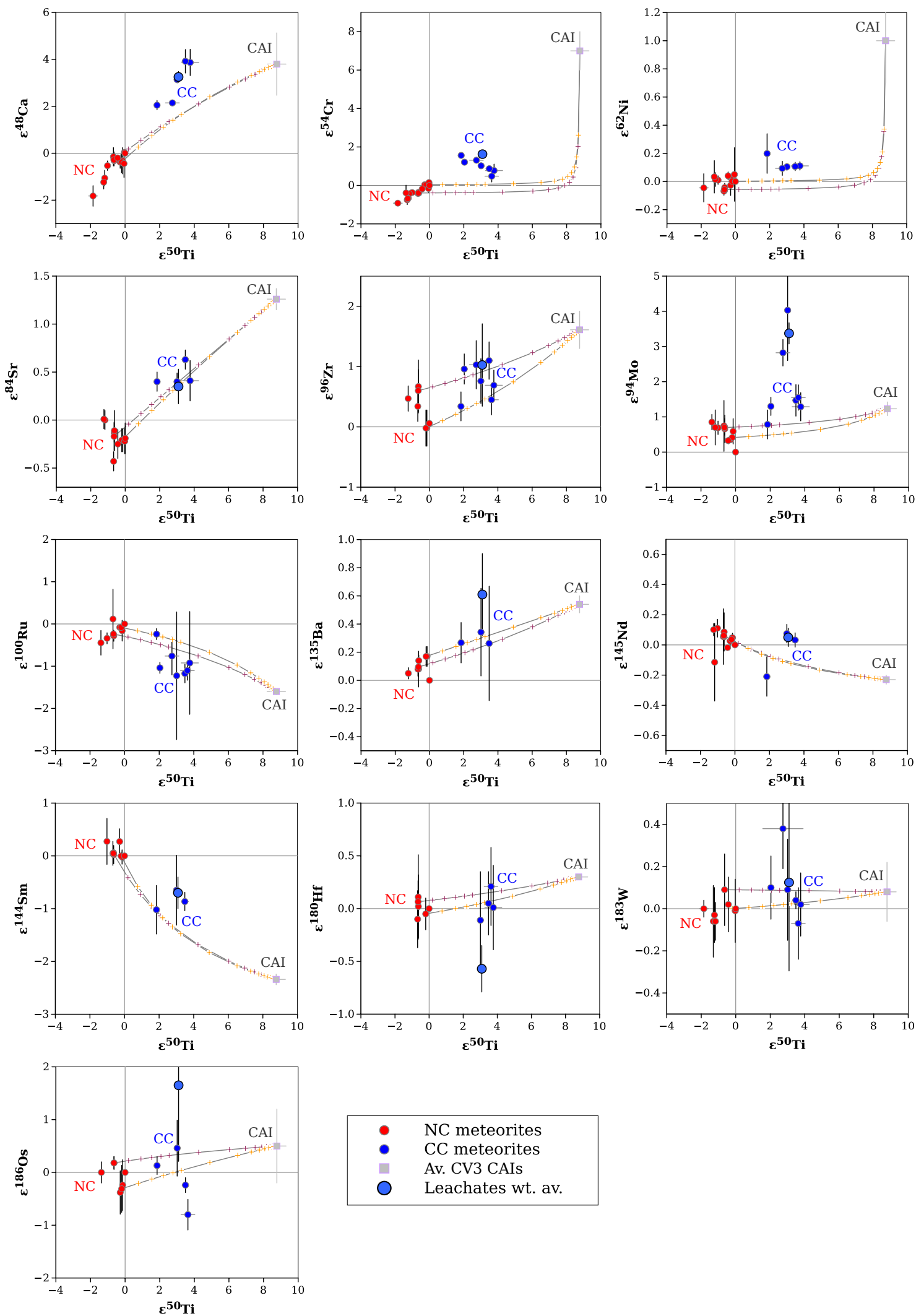

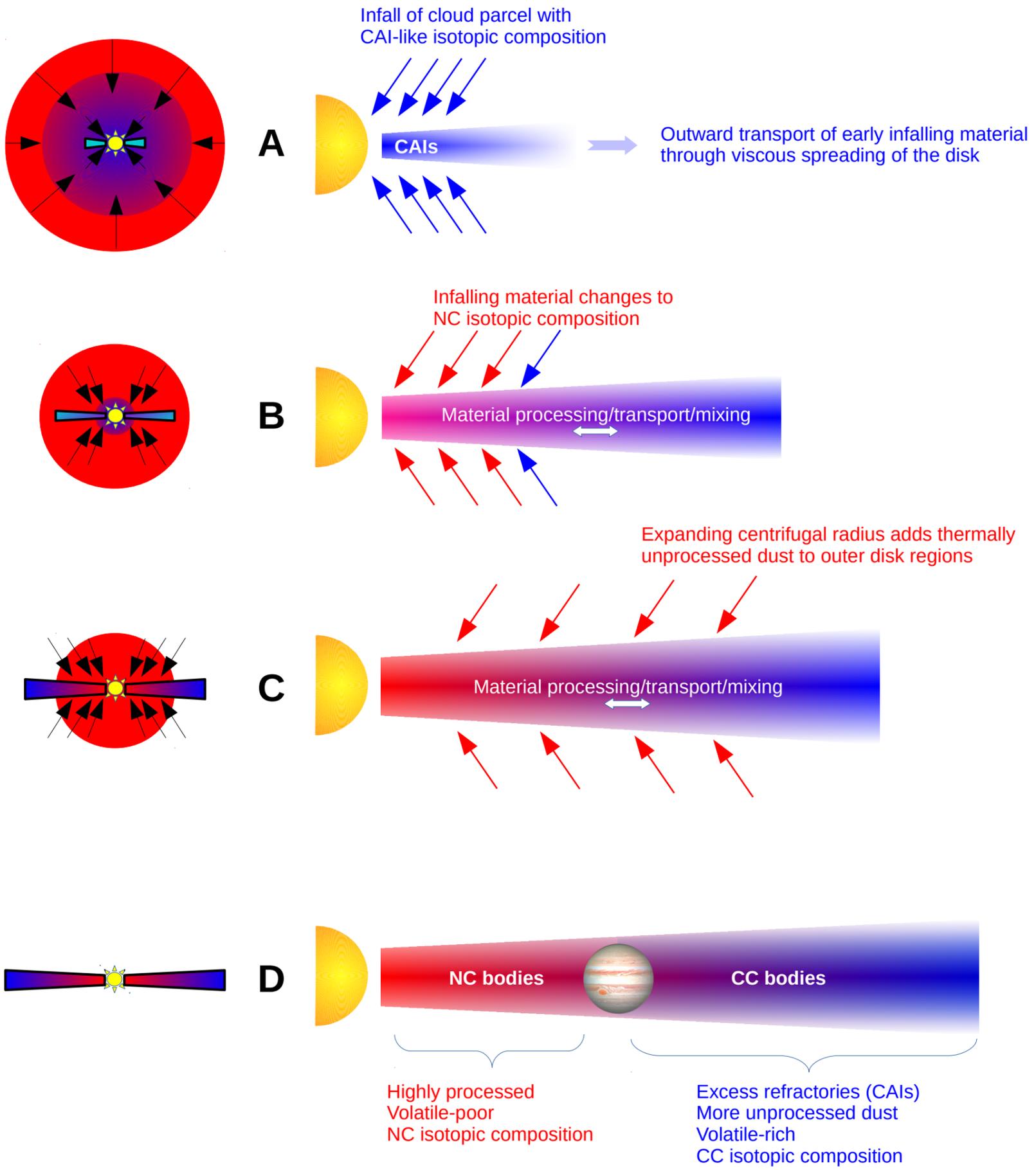

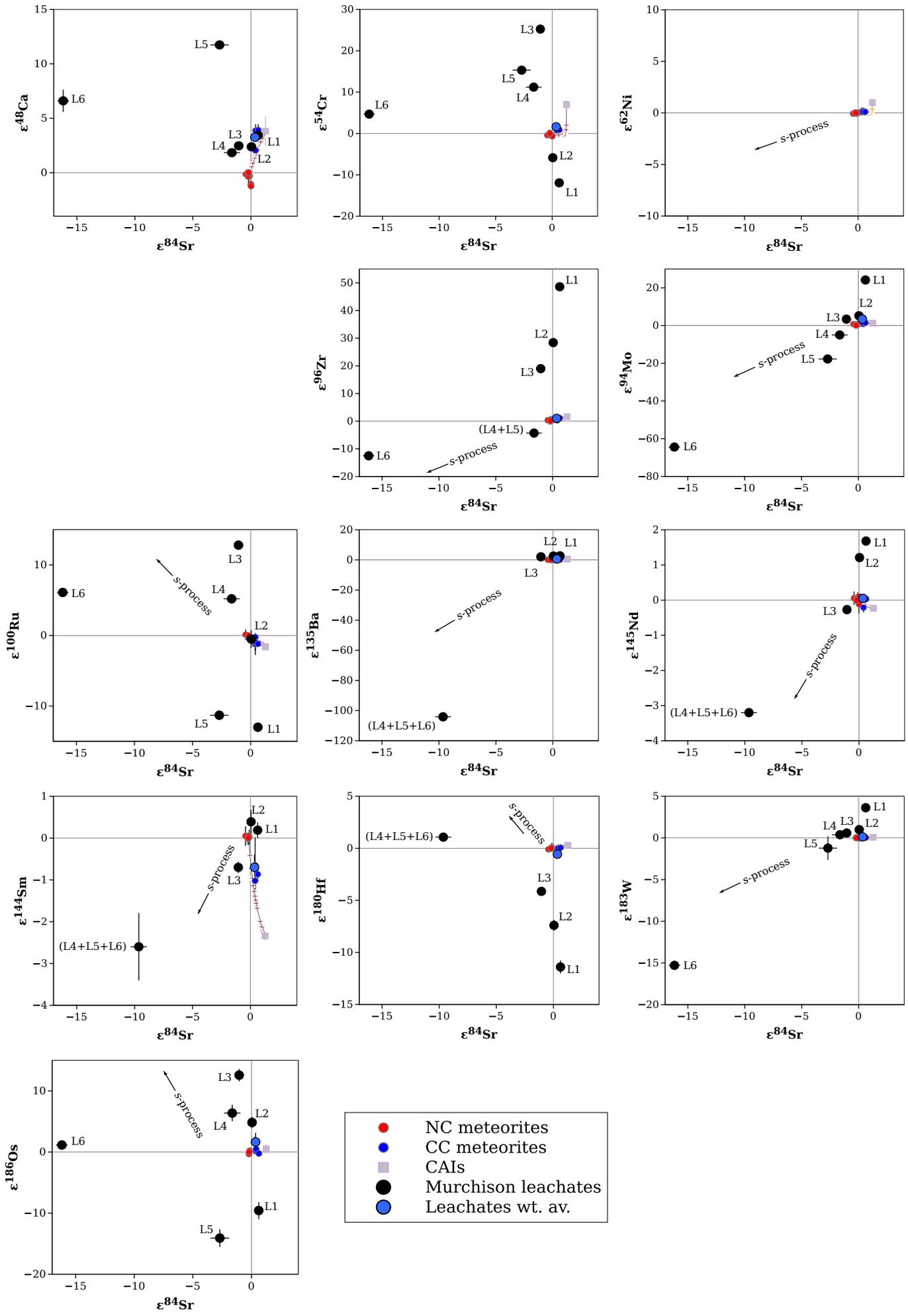

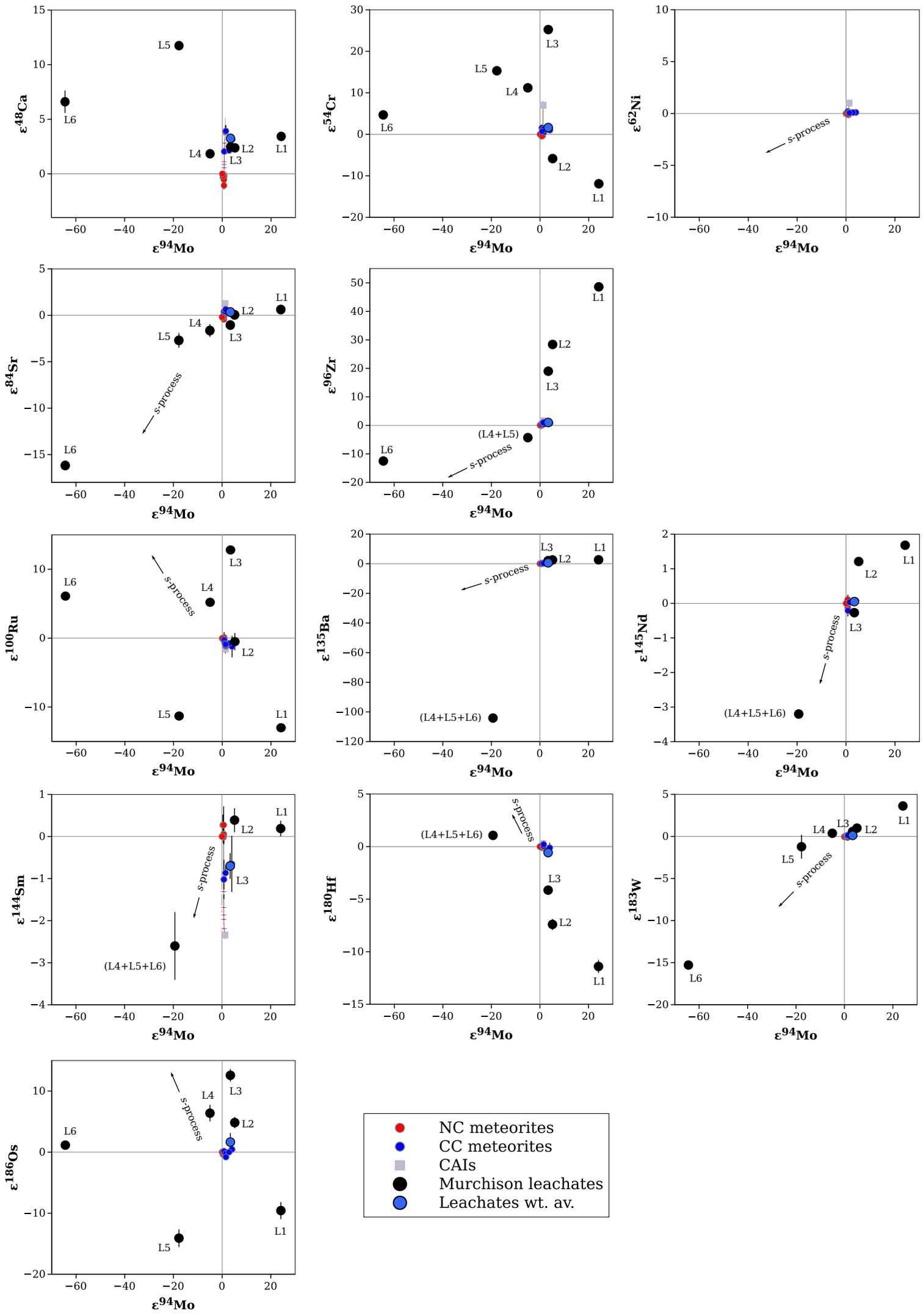

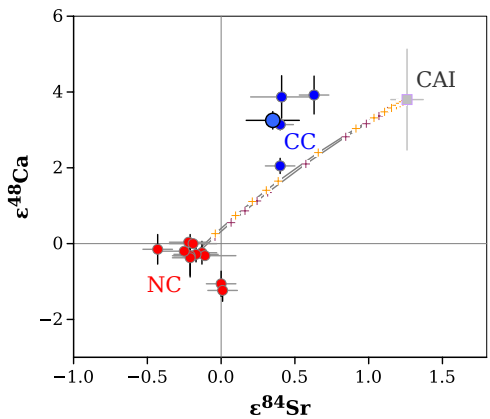
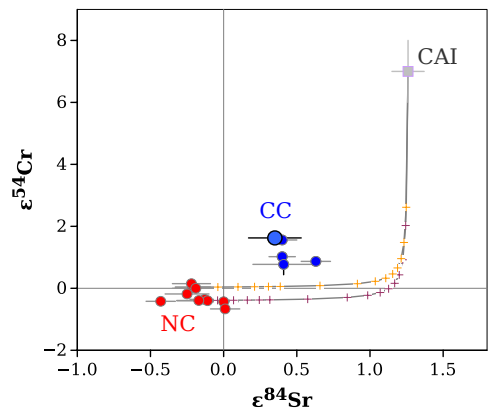
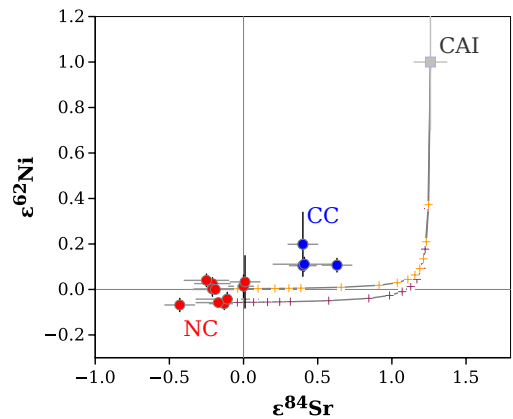
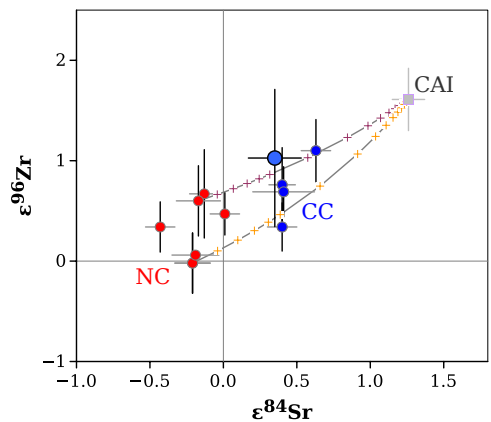
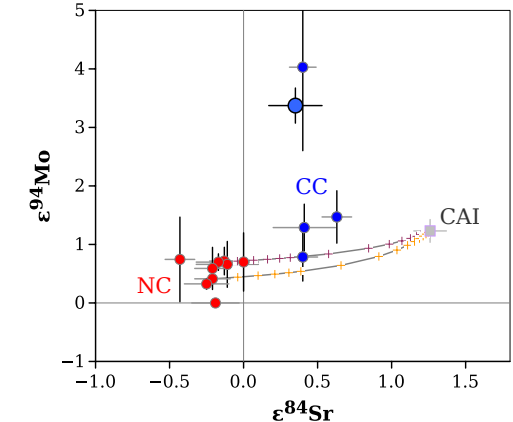
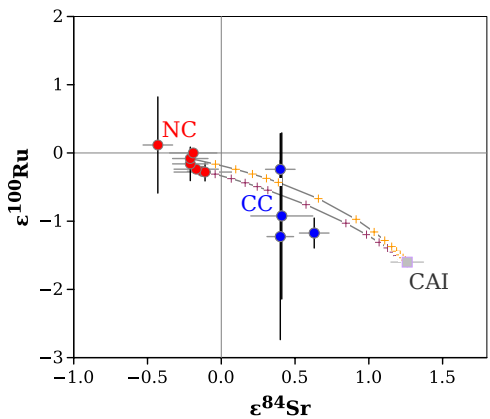
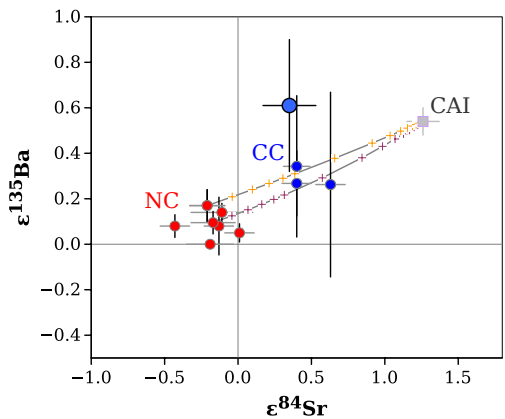
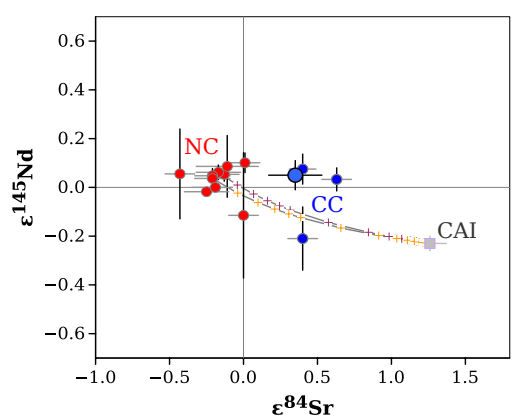
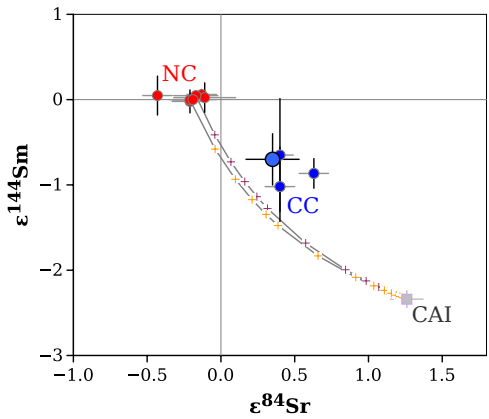
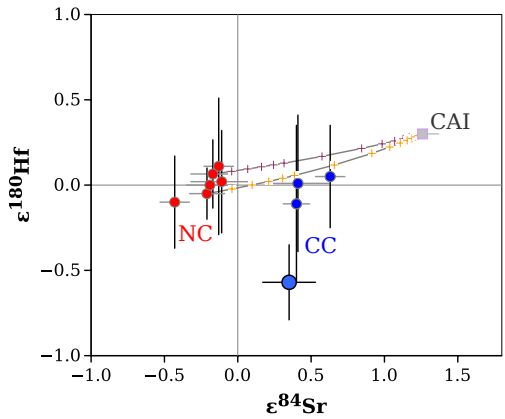
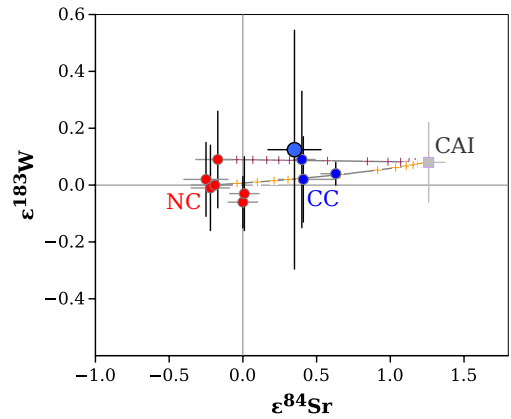
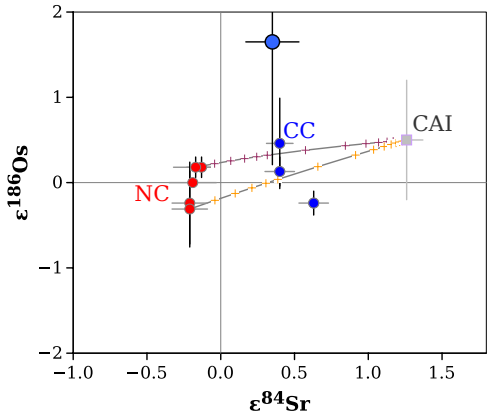
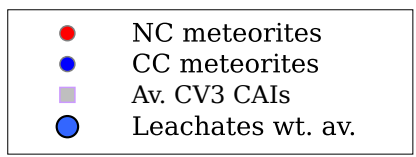

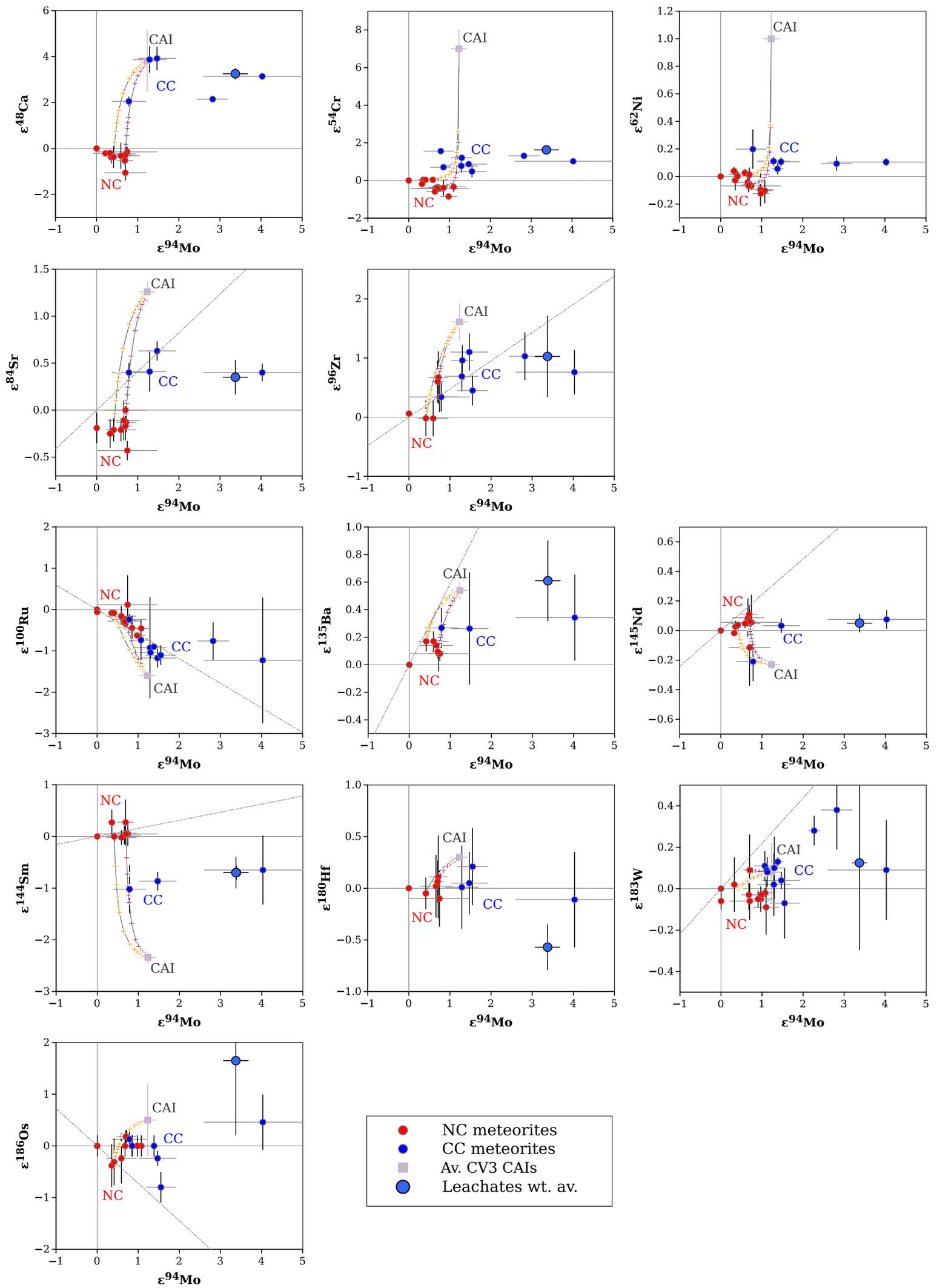